\documentclass[conference]{IEEEtran}
\IEEEoverridecommandlockouts
\usepackage{cite}
\usepackage{amsmath,amssymb,amsfonts}
\usepackage{tikz}
\usepackage{graphicx}
\usepackage{textcomp}
\usepackage{xcolor}
\usepackage{url}
\definecolor{mauve}{rgb}{0.58,0,0.82}
\def\BibTeX{{\rm B\kern-.05em{\sc i\kern-.025em b}\kern-.08em
    T\kern-.1667em\lower.7ex\hbox{E}\kern-.125emX}}
\makeatletter
\newcommand{\linebreakand}{%
  \end{@IEEEauthorhalign}
  \hfill\mbox{}\par
  \mbox{}\hfill\begin{@IEEEauthorhalign}
}
\makeatother
\DeclareMathAlphabet\mathbfcal{OMS}{cmsy}{b}{n}
\usepackage{scalerel}
\usepackage{xspace}
\usepackage{listings}
\usepackage{array}
\usepackage{pdfpages}
\usepackage{algorithm, algpseudocode}
\usepackage[table, dvipsnames]{xcolor}
\usepackage{makecell}
\usepackage{color}
\usepackage{colortbl}
\usepackage{bm}
\usepackage{fixmath} % for \mathbold
\usepackage{scalerel}
\newlength\bshft
\def\fakebold#1{\ThisStyle{\ooalign{$\SavedStyle#1$\cr%
  \kern-\bshft$\SavedStyle#1$\cr%
  \kern\bshft$\SavedStyle#1$}}}
\newcommand{\modelname}{$\textbf{MCAD}$\xspace}
\newcommand{\metricname}{$\protect{\textit{A}\textit{R}\textit{G}\textit{E}\textbf{-}\textit{A}\textit{D}}$\xspace}
\usepackage{multirow, makecell}
\usepackage{subfigure}
\usepackage{booktabs}
\usepackage{amsmath}
\usepackage{wrapfig}
\usepackage{amssymb}
\usepackage{multirow}
\usepackage{amsfonts}
\usepackage{enumitem}
\usepackage{tabularx}
\usepackage{bbm}
\usepackage{soul}
\usepackage{dsfont}
\usepackage{tablefootnote}
\usepackage{textcomp}
\usepackage{tabu,longtable}
\usepackage{refcount}      % Allows referencing counter values
\usepackage{afterpage}
\usepackage{lineno}
\usepackage{float}
\usepackage{bbold} % all ten digits with \mathbb
\usepackage{hhline}
\usepackage{pifont}
\newcommand{\shorteq}{%
  \settowidth{\@tempdima}{-}% Width of hyphen
  \resizebox{\@tempdima}{\height}{=}%
}

\newcolumntype{L}[1]{>{\raggedright\let\newline\\\arraybackslash\hspace{0pt}}m{#1}}
\newcolumntype{C}[1]{>{\centering\let\newline\\\arraybackslash\hspace{0pt}}m{#1}}
\newcolumntype{R}[1]{>{\raggedleft\let\newline\\\arraybackslash\hspace{0pt}}m{#1}}
\usepackage{threeparttable}
\usepackage{amsthm}

\begin{document}

\title{\modelname: Multimodal Context-Aware Audio Description Generation For Soccer}

% \author{\IEEEauthorblockN{1\textsuperscript{st} Lipisha Chaudhary}
% \IEEEauthorblockA{\textit{Department of Computer Science \& Engineering} \\
% \textit{University at Buffalo, SUNY}\\
% Buffalo, NY \\
% lipishan@buffalo.edu}

\author{\IEEEauthorblockN{Lipisha Chaudhary}
\IEEEauthorblockA{\textit{University at Buffalo, SUNY}\\
% Buffalo, NY \\
lipishan@buffalo.edu}
\and
\IEEEauthorblockN{Trisha Mittal}
\IEEEauthorblockA{\textit{Dolby Laboratories Inc.}\\
% City, Country \\
trisha.mittal@dolby.com}
\and
\IEEEauthorblockN{Subhadra Gopalakrishnan}
\IEEEauthorblockA{\textit{Dolby Laboratories Inc.}\\
% City, Country \\
subhadra.gopalakrishnan@dolby.com}
\linebreakand %
\IEEEauthorblockN{Ifeoma Nwogu}
\IEEEauthorblockA{\textit{University at Buffalo, SUNY}\\
% Buffalo, NY \\
inwogu@buffalo.edu}
\and
\IEEEauthorblockN{Jaclyn Pytlarz}
\IEEEauthorblockA{\textit{Dolby Laboratories Inc.}\\
% City, Country \\
jaclyn.pytlarz@dolby.com}
}

\maketitle

%%%%%%%%% ABSTRACT
\begin{abstract}
Audio Descriptions (AD) are essential for making visual content accessible to individuals with visual impairments. Recent works have shown a promising step towards automating AD, but they have been limited to describing high-quality movie content using human-annotated ground truth AD in the process. In this work, we present an end-to-end pipeline, \modelname, that extends AD generation beyond movies to the domain of sports, with a focus on soccer games, without relying on ground truth AD. To address the absence of domain-specific AD datasets, we fine-tune a Video Large Language Model on publicly available movie AD datasets so that it learns the narrative structure and conventions of AD. During inference, \modelname incorporates multimodal contextual cues such as player identities, soccer events/actions, and commentary from the game. These cues, combined with input prompts to the fine-tuned Video-LLM, allow the system to produce complete AD text for each video segment. We further introduce a new evaluation metric, \metricname, designed to accurately assess the quality of generated AD. \metricname evaluates the generated AD for the presence of five characteristics: (i) usage of people's names, (ii) mention of actions/events, (iii) appropriate length of AD, (iv) absence of pronouns, and (v) overlap from commentary/subtitles. We present an in-depth analysis of our approach on both movie and soccer datasets. We also validate the use of this metric to quantitatively comment on the quality of generated AD using our metric across domains. Additionally, we contribute audio descriptions for $100$ soccer game clips annotated by two AD experts.
% \footnote{We will release the code, our test sets and the AD expert annotations.}
\end{abstract}


% We propose MCAD, the first framework to generate automated Audio Descriptions (AD) for sports (soccer), extending AD beyond the movie domain. MCAD leverages video LLMs fine-tuned on movie AD and enriches them with contextual cues such as player identities, actions, and commentary. We further introduce ARGE-AD, a novel reference-free evaluation metric grounded in official AD guidelines. Experiments on movie and soccer datasets, plus expert-annotated clips, demonstrate MCAD’s ability to produce high-quality AD, moving towards inclusive and accessible sports broadcasting.

% In this work, we present an end-to-end pipeline~\modelname for enabling  AD generation for domains beyond movies, specifically sports (soccer games), without ground truth AD. 

% In this work, we present an end-to-end pipeline~\modelname for enabling  AD generation for domains beyond movies, specifically sports (soccer games), without ground truth AD. To workaround the unavailability of domain-specific AD datasets (outside of movies, like soccer games), we finetune a Video Large Language Model on the abundantly available movie AD datasets, to learn the structure of an AD for a given raw video. \modelname extracts essential multimodal contextual cues like player names, soccer events/actions, and commentary from the soccer game. This context is used along with input prompts to the finetuned Video-LLM to generate complete AD text as output.
\section{Introduction}
\label{sec:intro}
Audio Description (AD) is the descriptive spoken narration of visual content, primarily for assisting visual impairments in accessing visual content~\cite{acb}. In addition to aiding visually impaired audiences, AD also enhances media comprehension for autistic individuals, supports eye-free activities, facilitates child language development, and mitigates inattentional blindness for sighted users~\cite{lewis2021deep,perego2016gains}. AD allows individuals with low-vision access to descriptive audio tracks for various forms of visual experiences such as movies, live theater performances, television, museums, and sports. For live experiences such as sports and theater, these real-time ADs are delivered via earpieces or headphones provided by trained human describers using specialized microphones. For movies and television, descriptions are usually prerecorded and synced with the soundtrack, though live descriptions are sometimes used. 

\begin{figure}[t]
    \centering
    \includegraphics[width=0.75\columnwidth]{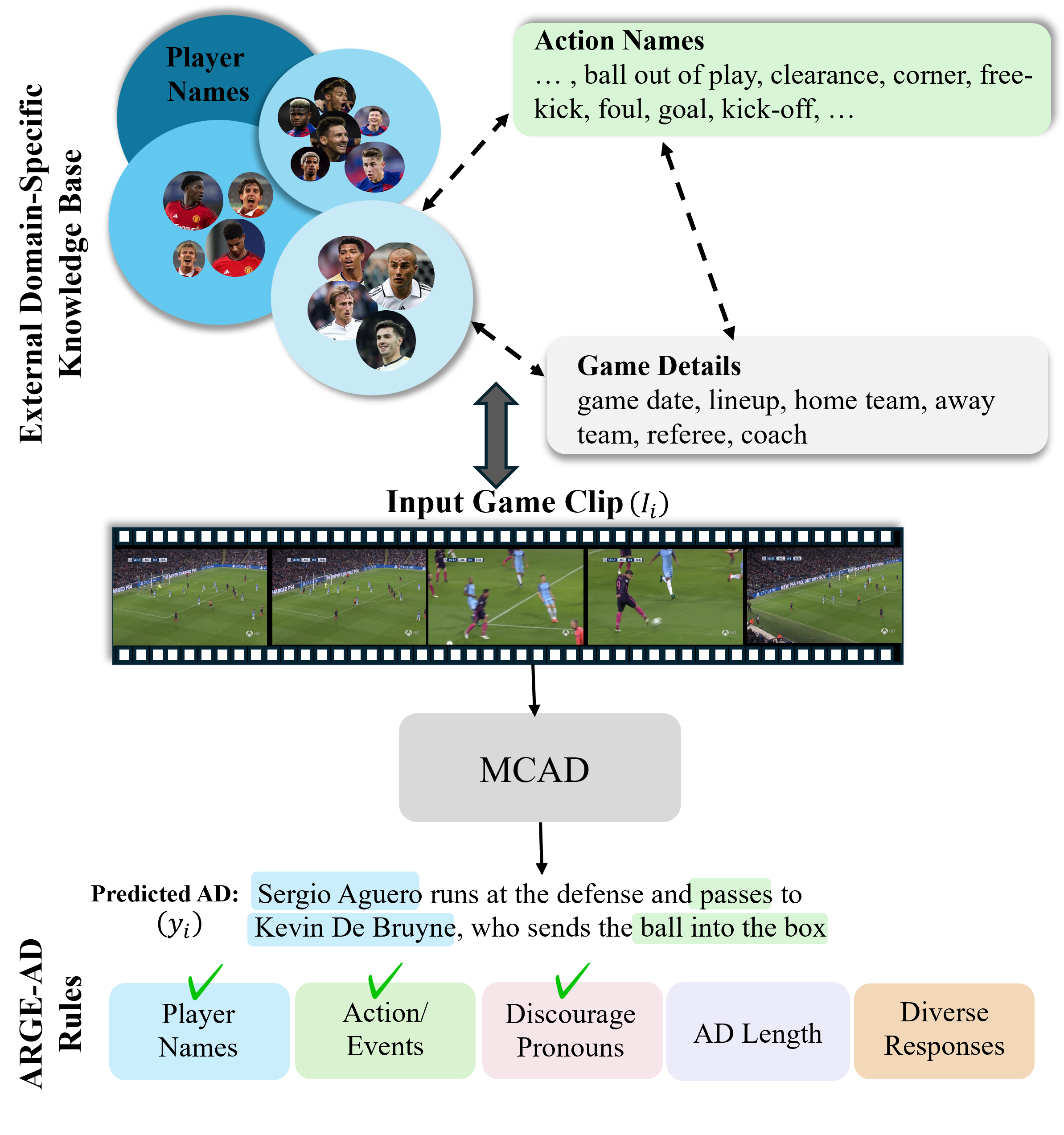}
    \caption{\small{\textbf{Autogenerating Audio Descriptions (ADs) For Soccer Games: } We propose \modelname, a framework to generate AD for domains beyond movies, with a focus on soccer games as a domain. \modelname enriches the generated AD by capturing all context cues like team, league, player names, actions, and commentary. We also propose \metricname, a reference-free metric based on AD conventions to evaluate the generated ADs.}}
    \label{fig:cover}
    \vspace{-15pt}
\end{figure}

Creating human-annotated ADs is costly and time-consuming, requiring skilled narrators to describe every visual detail in videos or live events. To address this, recent work has focused on automating AD generation using AI. Some prior multimodal models detect characters, actions, and context while capturing subtle cues and emotions~\cite{wang2021toward}. With advances in multimodal reasoning, Large Language Models (LLMs) and Visual Language Models (VLMs) have been used to generate ADs for movies~\cite{han2023autoad,han2023autoad2,xie2024autoad,zhang2024mm,han2024autoad,chu2024llm}. However, most efforts focus on movies, which offer structured scenes, fixed characters, and abundant annotated data. In contrast, sports broadcasting poses unique challenges—frequent camera shifts, crowd noise, overlapping commentary, and few natural pauses for AD insertion. Moreover, no domain-specific AD datasets exist for sports. In this work, we extend AD automation to soccer games without relying on domain-specific human annotations.

There is considerable overlap between AD and video captioning/description~\cite{aafaq2019video,krishna2017dense,li2021value,lin2022swinbert}, as both describe visual content. However, AD generation leverages multiple modalities to produce coherent narratives of storylines, characters, and actions that complement the existing audio. Organizations such as the American Council of the Blind (ACB)~\cite{brack_audio} and the DCMP~\cite{audio_description_tip_sheet} provide guidelines for AD composition—favoring simple, clear language that names characters and actions while avoiding pronouns, thus aligning with audio cues like commentary. Prior work~\cite{merullo2019investigating,AudioDescriptiveCommentary,AudioDescriptiveCommentary1} has shown that sports commentary often includes subjective bias, whereas AD, whether automated or manual, aims to remain objective.

% \textcolor{red}{Prior works~\cite{merullo2019investigating,AudioDescriptiveCommentary,AudioDescriptiveCommentary1} have shown commentator biases in sports broadcasts, which are intended to complement visual aids but inherently included subjective comments. Comparatively, generation of AD, automated or manual, retrain from adding biases or subjective opions.} 

It is also important to determine when an AD is needed based on gaps in the audio. However, existing metrics are based on traditional NLP measures~(which miss AD-specific aspects) or are reference-based. To address this, we propose a non-reference metric grounded in AD guidelines to quantitatively assess AD quality across domains.

\textbf{Main Contributions:} We make the following contributions- 
\begin{enumerate}
% \item First, we present \modelname, an AD generation framework for domains beyond movies, that are not based on supervised learning and hence are not limited to domains with the availability of rich, high-quality, human-annotated ground truth AD datasets. In this work, we focus on sports as our domain. 

 % Unlike traditional supervised learning approaches, MCAD eliminates the need for extensive labeled datasets. (trisha mittal: this sentence is redundant and can be deleted. )
 % The metric is based on the guidelines of creating a comprehensive AD in order to quantitatively comment on the quality of AD. This metric is not domain-specific and can be used for benchmarking AD generation methods across domains. 

 % \item Finally, a build a scaled-down dataset using expert AD annotators. We present this dataset to establish significant path in automating AD for face-paced sports videos. 

\item We introduce \modelname (Multimodal Context-Aware Audio Description), illustrated in Fig.~\ref{fig:cover}, a framework that automatically generates ADs for soccer games without relying on human-annotated training data, unlike prior models.

\item We also propose a novel evaluation metric, \metricname, to assess the overall quality of the generated AD.
% \metricname is a reference-free metric based on the official guidelines for creating a comprehensive AD as specified by ACB, DCMP, and other BLV advocacy organizations. 

\item We release a set of expert-annotated soccer clips, providing a first benchmark for AD in sports and supporting future research in this domain.

\end{enumerate}

We evaluate our approach and also validate our metric on the state-of-the-art movie AD datasets, CMD-AD~\cite{han2024autoad} and MAD-eval~\cite{han2023autoad}, as well as a soccer games dataset, SoccerNet-v2~\cite{giancola2018soccernet}. We also present $100$ soccer clips with annotations from AD experts, to illustrate the efficacy of \metricname. We also present some in-the-wild experiments for another sport, basketball, and another domain beyond sports and movies, real-life navigation.

\section{Related work}
\label{sec:related-work}

In this section, we review prior work in related domains. Section~\ref{subsec:impact-ad} discusses existing literature on the development and adoption of AD. Section~\ref{subsec:auto-ad-movies} covers key datasets and algorithms for automating ADs in movies. Finally, Section~\ref{subsec:auto-ad-sports} highlights prior efforts on AD generation beyond the movie domain.

\subsection{Existing Status of Audio Description}
\label{subsec:impact-ad}
ADs make videos accessible to blind and visually impaired individuals by translating visual content into audio. However, few videos include ADs due to the high cost and limited availability of professional AD services~\cite{thompson2017audio}, making them unaffordable for many casual creators. The challenge is even greater for live applications~(theater, sports, museums, real-time navigation, etc.).

\begin{figure*}[t]
    \centering
    \includegraphics[width=0.85\textwidth]{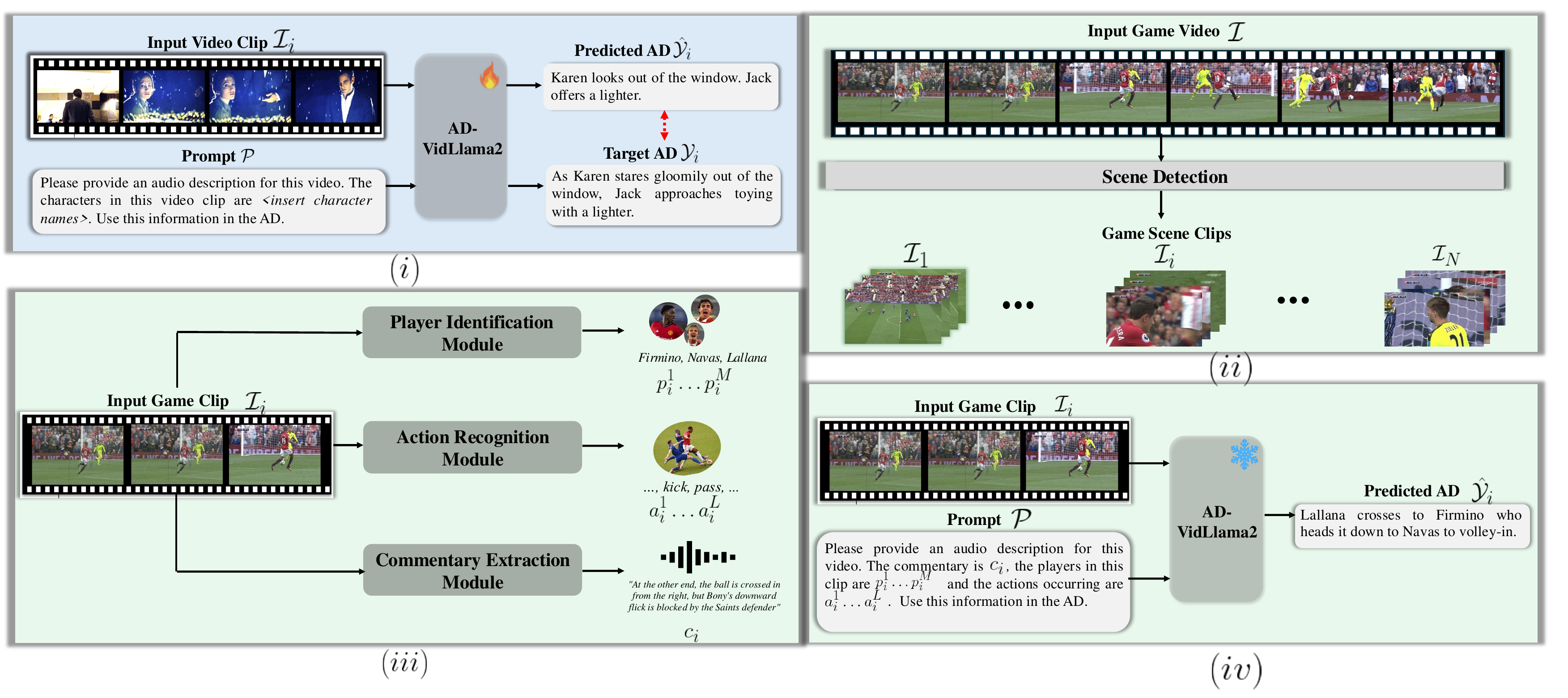}
    \caption{\small{\textbf{\modelname for Soccer Games: }We present an overview of our framework for using \modelname to generate AD for sports. The first step, ($(i)$) is the finetuning step, where we leverage the huge amount of available groundtruth AD data for movie clips, to develop \textit{AD-VidLlama2}, a Video-LLM that is enriched with AD aspects. In $(ii), (iii), (iv)$ we explain how we can perform an inference to now use the finetuned \textit{AD-VidLlama2} for generating ADs for sports game clips. In (ii) we take an entire game video, $\mathcal{I}$ and use scene detection to divide it into smaller game clips, $\mathcal{I}_1 \dots \mathcal{I}_i \dots \mathcal{I}_N$. In (iii) we focus on retrieving contextual cues for a particular game clip, $\mathcal{I}_i$. We get the corresponding commentary text $c_i$, player names $p_i^k$ and also actions $a_i^k$. And, finally in (iv) we combine the retrieved context in the prompt, $\mathcal{P}$ and the input video clip $\mathcal{I}_i$ to generate the AD $\mathcal{\widehat{Y}}_i$. }}
    \label{fig:overview-inference}
    \vspace{-15pt}
    
\end{figure*}

\subsection{Automated Audio Description for Movies}
\label{subsec:auto-ad-movies}

Efforts to automate AD generation for movies and TV shows have increasingly reduced dependence on manual processes, with early methods building on dense video captioning and multimodal deep learning~\cite{wang2021toward}. Recently, large language models (LLMs) such as PaLM~\cite{chowdhery2022palm} and BERT~\cite{devlin2018bert}, and visual language models (VLMs) like Video-LLaMA~\cite{liu2023video}, VideoBLIP~\cite{li2023videoblip}, MovieChat~\cite{fu2023moviechat}, GPT-4V~\cite{openai2023gpt4v}, and Flamingo~\cite{alayrac2022flamingo}, have shown strong multimodal reasoning and video understanding capabilities. These models are increasingly used as training-free solutions for AD automation. For instance, Han et al.~\cite{han2023autoad,han2023autoad2,han2024autoad} adapted GPT-2 for multimodal inputs via AD-specific adaptation layers, though their character recognition module required specialized training data. MM-VID~\cite{2023mmvid} combines multimodal perception experts with GPT-4V for video understanding, while MM-Narrator~\cite{zhang2024mm} uses a two-stage pipeline—GPT-4V for frame captions refined by GPT-4 into complete ADs, albeit without explicit character recognition. Such models benefit from rich movie datasets with AD annotations, including MAD~\cite{soldan2022mad}, CMD-AD~\cite{han2024autoad}, AudioVault-8k~\cite{han2023autoad}, and HowTo-AD~\cite{han2024autoad}.

% \subsection{Automated Audio Description For Sports}
\subsection{Audio Description in Multimedia}
\label{subsec:auto-ad-sports}
Although AD helps visually impaired individuals access visual media such as films, theater, and exhibitions~\cite{snyder2005audio}, limited research extends beyond movies. Sports, for instance, are the fifth-most popular TV genre among visually impaired viewers but the third-most difficult to follow due to their fast pace~\cite{pettitt1996audetel}. The Center for Access to Football in Europe (CAFE) provided audio descriptive commentary during UEFA EURO~2020~\cite{cafe2021}, and Japan’s NHK developed a real-time AD system for the Rio Olympics and Paralympics~\cite{kurihara2019automatic}. Despite such efforts and FCC regulations mandating 87.5 hours of audio-described programming per quarter~\cite{fcc2021audio}, most AD research and production remain focused on dramas, documentaries, lifestyle, and children’s content, with little work on automating AD beyond movies.

% \textcolor{red}{Similarly, Japan Broadcasting Corporation~(NHK) developed an automatic AD system for sports programs during the Rio Olympics and Paralympics~\cite{kurihara2019automatic}, though it has not yet been implemented in real-world scenarios}. % Despite international efforts and FCC regulations mandating $87.5$ hours of audio-described programming per calendar quarter~\cite{fcc2021audio}, much of the focus remains on drama, documentary, lifestyle, and children's content. 
% There is still a lack of prior research in automating AD in domains beyond movie content. 

% This is largely fueled by the lack of human-annotated AD dataset for sports and other applications. 

\section{Our Approach: \modelname}
\label{sec:approach}

A major challenge to generating audio descriptions (ADs) beyond movies is the lack of groundtruth data. We present an approach to extend Automated AD to general domains beyond movies, specifically focusing on Automated AD generation in soccer. Given a soccer game video, our goal is to generate ADs for all the scenes in that video. Our algorithm consists of a finetuning step followed by a context-aware inference model. Our overall approach is depicted in Fig~\ref{fig:overview-inference} and described in Algorithm \ref{alg:ad-generation} in Appendix~\ref{sec:appendix-a}.
% and in this paper, we demonstrate AD generation in sports videos.

\subsection{Finetuning to Develop an AD-rich Video-LLM}
\label{subsec:finetuning}

% We propose a Video Large Language Model (VideoLlama2)~\cite{damonlpsg2024videollama2} based model, ~\textit{AD-VidLlama2}, which effectively captures the intricate spatial and temporal dynamics of input video data. This process is illustrated in Figure~\ref{fig:overview-inference}-$(i)$. 

We propose \textit{AD-VidLlama2}, a Video-LLM~(VideoLlama2)~\cite{damonlpsg2024videollama2} fine-tuned on a movie AD dataset, CMD-AD \cite{han2023autoad}, to enhance the model's ability to process spatial and temporal dynamics of input videos while concurrently understanding aspects of AD. The process of finetuning is illustrated in Fig~\ref{fig:overview-inference}-$(i)$. VideoLlama2 employs a CLIP-based~\cite{pmlr-v139-radford21a} vision backbone that aids in easy incorporation of pixel-level video details. This helps the model seamlessly process intricate details in input videos during inference.

% In order to facilitate consistent feature extraction for improved video comprehension, the vision backbone records a set number of frames from every video. Before extracting spatial-temporal representations, the pre-trained VideoLlama2 pre-processes each of the eight video frames it receives. 
% ===========>
% The system extracts features uniformly by sampling a fixed number of frames from each video. Specifically, VideoLlama2's vision backbone pre-processes eight frames per video to generate consistent spatial-temporal representations. We test the number of input frames to see how it affects the production of AD.  

% model ensures that it learns and understands AD structure without giving long prompt instructions. Customized prompt structures are inherently subjective and heavily rely on factual knowledge\cite{wan2024efficient}. 
For fine-tuning~\textit{AD-VidLlama2}, we utilize the publicly available movie AD dataset, CMD-AD (details in Section \ref{subsec:ad-datasets}). Each movie video clip $\mathcal{I}_j, j \in [1,M]$, where $M$ is the total number of movie clips, is associated with an AD label $AD_{j}$. The \textit{AD-VidLlama2} uses the instruction prompt and movie clip $I_{j}$ as the input response. Specifically, we use the prompt \textit{``Give an audio description of the given video."} for finetuning. We use this movie-tuned \textit{AD-VidLlama2} model for generating automated AD in sports videos. 

% The pair ($I_{j}, AD_{j}$) is used as visual input with AD as the reference output. 

 % To generalize the framework, we adopt to instruction-tune existing established LLM.
% Following standard domain-adaptation techniques, we provide as input a dataset $\mathcal{D}$ consisting of $M$ video clips $\mathcal{ I}_j, j \in [1,M]$ and prompt $\mathcal{P}$, \textit{``Give an audio description of the given video''}, to a pretrained VLM. Each clip $\mathcal{I}_j$ has an associated groundtruth AD, $\mathcal{ Y}_j$. The fine-tuning objective is to minimize the cross-entropy loss $\mathcal{L}$  of the text tokens. 
% We use loss function $\mathcal{L}$ to finetune the VLM:

% \begin{equation}
%     \text{AD-VLM} \gets \mathcal{L} \left (\text{VLM}, \left\{\mathcal{ I}_{1:M}\right\}, \mathcal{P}\right )
%     \label{eq: finetuning}
% \end{equation}

% For each $\mathcal{I}_j$, we empirically observed that sampling $8$ frames yielded the optimal performance. Our empirical results further suggest that fine-tuning on any available video and AD pair, enable us to adapt it to a new domain. We use visual prompting to explicitly instruct the LLM to follow AD-specific rules to generate AD for any given video. Visual Prompting techniques are used to adapt large pre-trained language models to customized downstream tasks \cite{jia2022vpt}. Our work is similar to visual prompting which provides efficient ways to use a pre-trained VLM for task-specific fine-tuning .These observations align with \cite{damonlpsg2024videollama2}.

\subsection{AD Generation for Sports via \modelname}
\label{subsec:ad-gen-sports}

Audio description (AD) generation faces unique challenges when applied to domains like sports. While movies offer natural pauses between dialogues for inserting AD, sports events present three key technical challenges:

{
\begin{itemize}[noitemsep]
    \item Rapid event sequences, frequent camera switches, and fast player movements in sports require real-time AD processing.  
    \item Continuous crowd noise and commentary create a dense audio stream to be analyzed alongside visual data.  
    \item Unlike movies, sports broadcasts rarely have natural pauses, making insertion of AD without disrupting flow a challenge.  
\end{itemize}
}

% \end{enumerate}
To address these challenges, we introduce the \textit{Multimodal Context-Aware Audio Description (MCAD)} framework. MCAD processes two parallel streams - live video, ongoing commentary/ambient audio - to generate contextually appropriate audio descriptions (AD) in real-time. We leverage the capabilities of Video-based multimodal LLMs that generate relevant texts outputs by integrating visual features and corresponding contextual information. Fig~\ref{fig:overview-inference} shows an overview of the components of \modelname. 

% \textcolor{red}{Because the framework employs a large multimodal LLM architecture integrating visual features, audio signals, and contextual information, it results in the generation of timely and relevant AD.}

Our processing pipeline begins with temporal segmentation of a video of a soccer game half~\cite{soccer_info} $\mathcal{I}$ into M sequential clips ${\mathcal{I}_1, \mathcal{I}_2, ..., \mathcal{I}_M}$, shown in Figure~\ref{fig:overview-inference}-$(ii)$. For each clip $\mathcal{I}_i \in [1, M]$, the system generates a corresponding text-based ADs $\mathcal{Y}_i$. The segmentation algorithm, based on the approach in \cite{pyscenedetect}, detects scene transitions by analyzing frame-to-frame changes in pixel characteristics (color and intensity) against an empirically chosen threshold value. We generate logical scenes from longer soccer videos to mimic the duration of short clips from the movie datasets for fair comparisons. The soccer logical scenes are segmented based on scene cuts, actions, and major changes in scenarios, rather than using hard-coded scenes of fixed duration.

% \textcolor{red}{We do so to create logical scenes to replicate the short scene clips from movie datasets which create short clips based on scene cuts, actions and scenarios, instead of creating hard-coded scenes of fixed duration.}

\subsubsection{Retrieving Contextual Cues}
\label{subsec:prompting}
% The need for adding context in generating automated AD is crucial, which we add in the next step (Figure~\ref{fig:overview-inference}-$(iii)$). The three most critical pieces of context that add the most information to an AD is the information of the player, the action that player is performing in that clip, if any and the supporting commentary for the video clip. In this step, we retrieve the three contexts for the video clip. Each $\mathcal{I}_i$ is paired with a commentary $c_i$ obtained using off-the-shelf multi-lingual Automatic Speech Recognition (ASR) models \cite{10.5555/3618408.3619590}. 

% The player identification algorithm~\cite{Koshkina_2024_CVPR} uses a pose keypoint detector \cite{xu2022vitpose} to localize the player's upper body to capture the jersey number~\cite{Cioppa2022Scaling}. We employ a pretrained Scene-Text Recognition (STR) model based on \cite{bautista2022parseq} to recognize the jersey number from the cropped player's upper body. A standard STR algorithm automatically identifies and localizes text with images or videos. Our player identification method follows an Transformer based encoder-decoder architecture \cite{bautista2022parseq} with a trained language model. Each player $j$ identified in $\mathcal{I}_i$ is denoted as $p^j_i$.

% The overall player identification system \cite{bautista2022parseq} uses a Transformer-based encoder-decoder architecture enhanced with a language model for robust number recognitio

% (A) = $\{\text{team_name}\: (\text{jersey}: \text{player name}), ...\}$
Contextual integration is a key component of our AD generation pipeline (Fig.~\ref{fig:overview-inference}-$(iii)$). For each video segment $\mathcal{I}_i$, the system extracts three contextual elements: player identification, action classification, and commentary transcription. Commentary $c_i$ is obtained via multilingual ASR using Whisper~\cite{10.5555/3618408.3619590}. Player identification follows a multi-stage process~\cite{Koshkina_2024_CVPR}: a pose-based keypoint detector~\cite{xu2022vitpose} localizes upper bodies, and a pretrained Scene-Text Recognition (STR) model~\cite{bautista2022parseq} extracts jersey numbers. Player names are then retrieved from a curated mapping of jersey numbers to players, built from SoccerNet-v2 annotations. For each segment $\mathcal{I}_i$, detected players are denoted as $p_i^j$, where $j$ indexes individual players within $\mathcal{I}_i$.

% We introduce an action recognition model~\cite{precisespotting_eccv22} into our AD generation pipeline to identify soccer specific events in the input clips. For training of the action recognition module, we use benchmark \cite{Deliège2020SoccerNetv2} dataset to detect temporally spanned soccer-related events in a given clip. The architecture uses a standard Temporal Action Detection (TAD) algorithm to localize and detect an action. The architecture uses RegNet-Y \cite{Radosavovic_2020_CVPR} CNN as the 2D backbone and GRU \cite{69e088c8129341ac89810907fe6b1bfe} for capturing long-term temporal context in the sports clip. Specifically, we detect soccer actions like ``Ball out of play'',  ``Corner'', ``free-kick'', ``Foul'', ``Goal", ``Yellow card'', ``Red card'', etc. Each action $k$ identified in $\mathcal{I}_i$ is denoted as $a^k_i$.

Our system integrates an action recognition model~\cite{precisespotting_eccv22} to automatically detect soccer events within video clips. Trained on the SoccerNet-v2 dataset~\cite{Deliège2020SoccerNetv2}, which provides labeled footage with precise event timestamps, it uses a Temporal Action Detection (TAD) approach to locate and classify actions as they occur. The architecture combines a RegNet-Y CNN~\cite{Radosavovic_2020_CVPR} for frame-level analysis with a GRU~\cite{69e088c8129341ac89810907fe6b1bfe} for temporal modeling. This enables detection of key events such as \texttt{Corner}, \texttt{Free-kick}, \texttt{Foul}, \texttt{Goal}, \texttt{Yellow card}, \texttt{Red card}, and \texttt{Ball out of play}, represented as $a_i^k$ for each action $k$ in clip $\mathcal{I}_i$.
Temporal segmentation and action recognition synchronize AD timing with real game events, ensuring each description aligns with a coherent soccer moment. To better reflect match context, the system jointly processes visual cues and live commentary, encoding video frames alongside ASR-transcribed speech. This fused representation grounds the generated ADs in both visual and linguistic context, resulting in more accurate and natural descriptions.

\subsubsection{Context-rich \textit{AD-VidLlama2} Prompting}
% In the final step, we feed in the input video clip $\mathcal{I}_i$, along with the prompt $\mathcal{P}$, that includes the three context cues, transcribed commentary $p_i$, identified player $p^j_i$ and, detected soccer action $a^k_i$ into ~\textit{AD-VidLlama2} using text prompting.

In the final step, we provide the \textit{AD-VidLlama2} model with the input video $\mathcal{I}_i$ and the prompt $\mathcal{P}$, which incorporates three contextual cues: transcribed commentary $c_i$, identified players $p_i^j$, and detected actions $a_i^k$, as shown in Fig.~\ref{fig:overview-inference}-$(iv)$.
{\small
\begin{equation}
\mathcal{Y}_i = \text{\textit{AD-VidLlama2}}\big(\mathcal{I}_i, c_i, p_i^j, a_i^k, \mathcal{P}\big)
\label{eq:frozen_inference}
\end{equation}}
Providing this contextual information enables \textit{AD-VidLlama2} to enrich AD generation with domain-specific insights. Details of the prompt design $\mathcal{P}$ are included in Appendix~\ref{sec:appendix-a}.

\section{Dataset, Metric and Experiment Details}
\label{sec:implementation-details}

Section~\ref{subsec:ad-datasets} describes the datasets used to evaluate our method, and Section~\ref{subsec:ad-metrics} outlines traditional text-matching metrics for AD evaluation. Section~\ref{subsubsec:arge} introduces \metricname, our new AD evaluation metric.

% \subsection{AD Evaluation Datasets}
\subsection{Datasets}
\label{subsec:ad-datasets}
We use SoccerNet-v2 dataset~\cite{Deliège2020SoccerNetv2} for our sports use-case. This dataset contains $500$ complete European Soccer League games, with each match following the standard format of two $45$-minute halves. SoccerNet-v2 provides comprehensive game-related annotations including major game events and detailed player information, details given in the Appendix \ref{sec:appendix-a}.
For our evaluation, we use randomly sampled $\sim2000$ clips from the test set of $100$ games, to mimic the sample size fo the test sets in the movie datasets. We refer to this curated dataset as \textbf{Soccernet-S}. To divide this into clips, we use a scene detection algorithm as described in Section~\ref{subsec:ad-gen-sports}, generating clips roughly $15$s to $40$s in length.

Our sports Auto-AD framework leverages two distinct movie datasets: CMD-AD \cite{han2023autoad} for training and initial testing, and MAD-Eval \cite{han2023autoad} for comprehensive evaluation. The CMD-AD dataset contains key scenes from movies, extracted as short clips from YouTube. These clips were specifically selected to capture important story moments, resulting in~$86,304$ training samples and $6,198$ test samples.\footnote{Due to the unavailability of a few video clips, our training and test size do not match the actual test and train set sizes, The size of original dataset: $93,951$ train samples and $7,316$ test samples}
The MAD-Eval dataset~\cite{han2024autoad} complements this with $10$ movies, sourced from the LSMDC~\cite{lsmdc} validation and test sets.  We specifically used the \textit{"Unnamed"} version of MAD-Eval. Both these datasets contain short clips, ranging from 30 seconds to 2 minutes.

\subsection{Evaluation Metrics}
\label{subsec:ad-metrics}
% Below, we discuss metrics that have been traditionally used by prior work to evaluate the quality of generated AD and also elaborate on our proposed metric.

The following section examines both traditional AD evaluation metrics from prior work and presents our proposed evaluation metric \metricname.
% qual-img.png
%%%%%%%%%%%%%%%%%%%%%%%%%
\begin{figure*}[h]
    \centering
    \includegraphics[width = 0.85\textwidth]{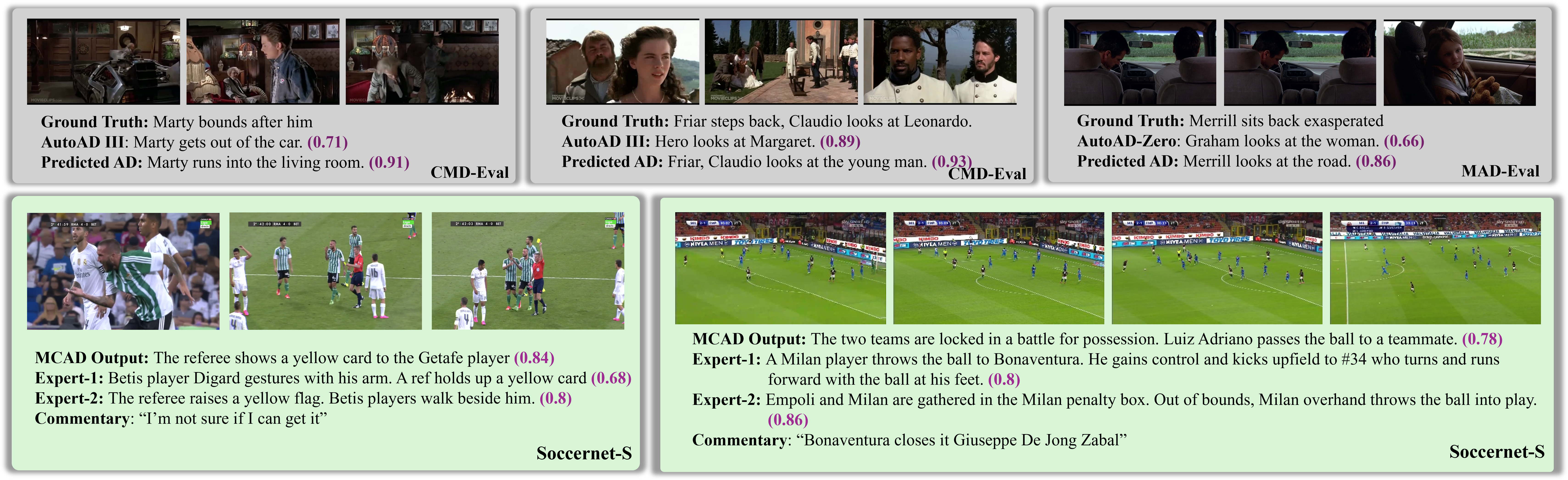}
    \caption{\small{\textbf{Qualitative Results: }We show qualitative visualizations for CMD-AD (top) and SoccerNet (bottom). The examples are from \textit{``The Man Who Wasn't There" \textbf{(top left)}}, \textit{``Back to the Future" \textbf{(top middle)}}, and \textit{``Much Ado About Nothing" \textbf{(top right)}}. The SoccerNet-S examples are from \textit{Real Madrid vs Betis [Spain LaLiga (2015-2016)], scene length is 15 secs} \textbf{\textit{(bottom left)}}  and \textit{AC Milan vs Empoli [Italy Serie A (2015-2016)] \textbf{(bottom right) scene length is 30 secs.}}}} 
    \label{fig:qualitative-results-image}
    \vspace{-5pt}
\end{figure*}

\subsubsection{Existing Metrics Used for AD Evaluation}

\textbf{Classic metrics for text generation.} Following prior work on automated AD, we use CIDEr~\cite{vedantam2015cider} and ROUGE-L~\cite{lin-2004-rouge}, two standard captioning metrics for evaluation. CIDEr applies TF-IDF weighting to assess word relevance and distinctiveness, while ROUGE-L measures similarity based on the longest common subsequence. However, both metrics often fail to capture semantically equivalent descriptions expressed with different wording.

\textbf{Newer metrics used for AD.} We also employ CRITIC~\cite{han2024autoad} and LLM-AD-Eval~\cite{han2024autoad}, two metrics specifically designed for AD evaluation. CRITIC assesses how accurately characters are identified by using a co-reference model to replace ambiguous pronouns with official character names and then computing an IoU score between name sets. LLM-AD-Eval leverages large language models such as LLaMA2-7B-Chat~\cite{touvron2023llama} and LLaMA3-8B-Instruct~\cite{dubey2024llama} to rate ADs on a 1–5 scale based on their alignment with the ground truth in terms of actions, objects, and interactions.

%%%%%%%%%%%%%%%%%%%%%%%%%
% \begin{table*}[ht]
%     \centering
%     \resizebox{.7\textwidth}{!}{
%     \begin{tabular}{rcccccccc}
%     \toprule
%     \textbf{Models} & \multicolumn{5}{c}{\textbf{CMD-AD}} & \multicolumn{3}{c}{\textbf{MAD-Eval}}  \\
%     \cmidrule(rr){2-6} 
%     \cmidrule(ll){7-9}
%     & \textbf{CIDEr} & \textbf{ROGUE-L} & \textbf{CRITIC} & \textbf{LLM-AD-eval}  & \textbf{\metricname} & \textbf{CIDEr} & \textbf{ROGUE-L} & \textbf{\metricname} \\ \hline
%     AutoAD~\cite{han2023autoad} & - & - & - & - & - & 12.1 & 10.3 & \\ 
%     AutoAD-II~\cite{han2023autoad2} &  13.5 & - & 8.2 & 2.08 & - & 19.2 & 13.1 &  \\ 
%     AutoAD-III~\cite{han2024autoad} &  (\textit{25.0})  & & 32.7 & 2.85 & 40.92 & 24.0 &  &  \\ 
%     Uni-AD~\cite{wang2024contextual} &  - & - & - & - & - & 27.3  & 16.8  & \\ %\midrule 
%     MM-Narrator (GPT-4)~\cite{zhang2024mm} &  - & - & - & - & - & 13.9 & 13.4 &   \\ 
%     MM-Narrator (GPT-4v)~\cite{zhang2024mm} & - & - & - & - & - & 9.8  & 12.8 &  \\ 
%     LLM-AD~\cite{chu2024llm} &  - & - & - & - & - & 20.5  & 13.5 &   \\ 
%     AutoAD-Zero~\cite{xie2024autoad}  &  17.7 & - & 43.7 & 2.83 & - & 22.4  & - &  \\ \hline
%     \textbf{Ours} &  &  &  &  &  &  &  &  \\
%     \bottomrule
%     \end{tabular}
%     }
%     \caption{\textbf{Evaluation on AD for Movies: }Comparison of our method, with other methods on CMD-AD and MAD-Eval datasets using $5$ metrics: CIDEr, ROGUE-L, CRITIC, LLM-AD-eval and \metricname.}
%     \label{tab:comparison-movies}
% \end{table*}

\begin{table*}[ht]
    \centering
    \resizebox{.9\textwidth}{!}{
    \begin{tabular}{rcccccccc}
    \toprule
    \textbf{Models} & \multicolumn{5}{c}{\textbf{CMD-AD}} & \multicolumn{3}{c}{\textbf{MAD-Eval}}  \\
    \cmidrule(rr){2-6} 
    \cmidrule(ll){7-9}
    & \textbf{CIDEr $\uparrow$} & \textbf{ROGUE-L $\uparrow$} & \textbf{CRITIC $\uparrow$} & \textbf{LLM-AD-eval $\uparrow$}  & \textbf{\metricname $\uparrow$} & \textbf{CIDEr $\uparrow$} & \textbf{ROGUE-L $\uparrow$} & \textbf{\metricname $\uparrow$}\\ \hline
    AutoAD~\cite{han2023autoad} & - & - & - & - & - & 12.1 & 10.3 & -\\ 
    AutoAD-II~\cite{han2023autoad2} &  13.5 & - & 8.2 & 2.08 & - & 19.2 & 13.1 & - \\ 
    AutoAD-III~\cite{han2024autoad} &  20.4 \textit{(25.0)}  & 20.3 & 33.18 (33.36) & 1.58 (2.85) & 0.41 & 24.0 & - &  -\\ 
    Uni-AD~\cite{wang2024contextual} &  - & - & - & - & - & 27.3  & 16.8  & -\\ %\midrule 
    MM-Narrator (GPT-4)~\cite{zhang2024mm} &  - & - & - & - & - & 13.9 & 13.4 &  - \\ 
    MM-Narrator (GPT-4v)~\cite{zhang2024mm} & - & - & - & - & - & 9.8  & 12.8 & - \\ 
    LLM-AD~\cite{chu2024llm} &  - & - & - & - & - & 20.5  & 13.5 & - \\ 
    AutoAD-Zero~\cite{xie2024autoad}  &  13.97 (17.7) & 18.8 & 35.51 (43.7) & 1.32 (2.83) & 0.43 & 22.4  & - & - \\ \hline
    \textbf{\modelname (Ours)} & 24.9 & 23.0 & 65.85& 1.41 & 0.55 & 34.4 & 24.10 & 0.53 \\
    \bottomrule
    \end{tabular}
    }
    \caption{\textbf{Evaluation on AD for Movies: }Comparison of our method, with other methods on CMD-AD and MAD-Eval datasets using $5$ metrics: CIDEr, ROGUE-L, CRITIC, LLM-AD-eval and \metricname. For AutoAD-III~(row 3) and AutoAD-Zero~(row 9), values in brackets are those reported by the authors when evaluated on the entire test set; we also report values that we obtain when we test their provided inferences for our test set of $6,198$ clips. Note that all other methods except \modelname  (ours) require ground truth data for training.}
    \label{tab:comparison-movies}

\end{table*}

% ,0.37336486523788437,0.06666666666666667,1.6681042565160322,0.19787693076391608,0.14379618563534843

% ,experiment_name,arge4ad,critic_score,llm_ad_eval_score,Bleu_1,Bleu_2,Bleu_3,Bleu_4,ROUGE_L,CIDEr
% output-movie-finetuned-wcontext-cmd-prompt3_finetune_videollama2_vllava_04_10_16_prompt1_checkpoint-6000,0.36656182683558625,0.1616924820551568,1.2896095514682155,0.1732912380751021,0.0495334383593381,0.021027283488173884,0.010404770136994526,0.19352223412833272,0.15153389748395243

    % \textbf{Ours} & 13.36  & 9.15 & 1.21 & &  & & & \\

\begin{table}[t]
    \centering
    \resizebox{\columnwidth}{!}{
    \begin{tabular}{rcc}
    \toprule
    \textbf{Models} & Domain  & \textbf{\metricname $\uparrow$} \\ \hline
    SoccerNet-caption \cite{Mkhallati_2023_CVPR} & Video Captioning & 0.45   \\ 
    % GOAL \cite{6487e421a6d94945bb750362d0e6104a} &&  \\ 
    % DVC-1 \cite{cioppa2024soccernet2024challengesresults} & & \\ \midrule
    \modelname (\textbf{Ours}) & Audio Description Generation & 0.73\\
    \bottomrule
    \end{tabular}
    }
    \caption{\small{\textbf{Evaluation on AD for Sports: }Performance  of our method along with Dense Video Captioning annotations methods  reported on ARGE4AD on SoccerNet dataset. }}
    \label{tab:comparison-sports}
\end{table}
\begin{table*}[ht]
    \centering
    \resizebox{0.85\textwidth}{!}{
    \begin{tabular}{cccccccccccc}
        \toprule[1.5pt]
          \multicolumn{3}{c}{\textbf{\large Context}} & \multicolumn{5}{c}{\textbf{\large CMD-AD}}  & \multicolumn{3}{c}{\textbf{\large MAD-Eval}} & \textbf{\large Soceernet-S} \\
         \cmidrule(rr){1-3}
         \cmidrule(lr){4-8}
         \cmidrule(lr){9-11}
         \cmidrule(ll){12-12 }
         \makecell{\textbf{character/player $+$} \\ \textbf{action/event}} & \makecell{\textbf{subtitles/} \\ \textbf{commentary}}  & \makecell{\textbf{previous} \\\textbf{AD}} & \textbf{CIDEr}  & \textbf{ROGUE-L} & \textbf{CRITIC} & \textbf{LLM-AD-EVAL}  & \textbf{\metricname} & \textbf{CIDEr}  & \textbf{ROGUE-L} & \textbf{\metricname} & \textbf{\metricname} \\
        \midrule
        \cellcolor{red!25}\ding{55} & \cellcolor{red!25}\ding{55} & \cellcolor{red!25}\ding{55} & 12.3 & 18.7 & 2.82 & 1.14 & 0.34 & 14.15 & 20.30 & 0.36 & 0.69 \\
        \cellcolor{green!25}\checkmark  & \cellcolor{red!25}\ding{55} & \cellcolor{red!25}\ding{55} & 24.9 & 23.0 & 65.85 & 1.41 & 0.55 & 34.4 & 24.10 & 0.53 & 0.73\\
        \cellcolor{green!25}\checkmark& \cellcolor{green!25}\checkmark & \cellcolor{red!25}\ding{55} & - & - & - & - & - & - & - & - & 0.70\\
        % \cellcolor{green!25}\checkmark & \cellcolor{green!25}\checkmark & \cellcolor{red!25}\ding{55} & - & - & - & - & - & - & - & - & 41.87 \\
        % \cellcolor{green!25}\checkmark  & \cellcolor{red!25}\ding{55} & \cellcolor{green!25}\checkmark & - & - & - & - & - &  - & - & - & 0.65 \\
        \cellcolor{green!25}\checkmark  & \cellcolor{green!25}\checkmark & \cellcolor{green!25}\checkmark & - & - & - & - & - &  - & - & - & 0.69 \\
        \bottomrule[1.5pt]
    \end{tabular}}
    
    \caption{\small{\textbf{Ablations Experiments for Context: }We perform an ablation experiment with using different context information in the prompt during inference. We test it for both the movie datasets, CMD-AD and MAD-Eval and also the SoccerNet dataset. \textit{(Note we do not have information about subtitles and previous clip's AD for the movie datasets, hence the '-')}. For these experiments, we used Prompt 3 for inference for all three datasets, keeping number of frames at 16 for the movie datasets and 4 for Soccernet-S.  }}
    \label{tab:context-ablations}
    % \vspace{-15pt}
\end{table*}

\begin{table*}[ht]
    \centering
    \resizebox{.8\textwidth}{!}{
    \begin{tabular}{rccccccccc}
    \toprule[1.5pt]
    \textbf{Prompts} & \multicolumn{5}{c}{\textbf{CMD-AD}} & \multicolumn{3}{c}{\textbf{MAD-Eval}} & \multicolumn{1}{c}{\textbf{Soceernet-S}}  \\
    \cmidrule(rr){2-6} 
    \cmidrule(lr){7-9}
    \cmidrule(ll){10-10}
    & \textbf{CIDEr} & \textbf{ROGUE-L} & \textbf{CRITIC} & \textbf{LLM-AD-eval}  & \textbf{\metricname} & \textbf{CIDEr} & \textbf{ROGUE-L} & \textbf{\metricname} & \textbf{\metricname}\\ \hline
    Prompt 1 & 20.0 & 22.8 & 61.7 & 1.30 & 0.53 & 35.48 & 24.78 & 0.53 & 0.71\\ 
    Prompt 2 & 19.1 & 21.4 & 62.17 & 1.24 & 0.55 & 36.26 & 24.43 & 0.52 & 0.72\\ 
    Prompt 3 & 24.9 & 23.0 & 65.85 & 1.41  & 0.55 & 34.4 & 24.10 & 0.53 & 0.73\\ 
    \bottomrule[1.5pt]
    \end{tabular}
    }
    
    \caption{\small{\textbf{Ablations Experiments for Different Prompts: }We perform an ablation experiment with using three different prompts, referred to as Prompt 1, Prompt 2 and Prompt 3 here~(only during inference time). \textit{AD-VidLlama2} was finetuned using Prompt 1. We test it for both the movie datasets, CMD-AD and MAD-Eval and also the SoccerNet-S dataset. The number of frames used for inferencing per clips was 16 for CMD-AD and MAD-Eval datasets, and 4 for SoccerNet-S.}}
    \label{tab:prompt-ablations}
\end{table*}

\begin{table}[h!]
    \centering
    \resizebox{.77\columnwidth}{!}{
    \begin{tabular}{ccccc}
        \toprule
        \textbf{Ground Truth} & \multicolumn{2}{c}{\textbf{CIDEr}} & \multicolumn{2}{c}{\textbf{ROUGE-L}} \\ 
        \cmidrule(lr){2-3}
        \cmidrule(lr){4-5}
         & \textbf{Pearson} & \textbf{Kendall} & \textbf{Pearson} & \textbf{Kendall} \\ 
        \midrule
        Annotator 1 & \multirow{2}{*}{0.84} & \multirow{2}{*}{0.37} & \multirow{2}{*}{0.57} & \multirow{2}{*}{0.32} \\ 
        Annotator 2 &  &   &  &  \\ 
        \bottomrule
    \end{tabular}}
\caption{\small{\textbf{Comparing \modelname against Human Annotations: }We benchmark the performance of \modelname against the annotations obtained from AD experts for $107$ Soccernet video clips.}}
\label{tab:human-annotation}
\vspace{-15pt}
\end{table}

% % \kappa
% % https://www.statology.org/cohens-kappa-python/

% \begin{table}[ht]
%     \centering
%     \resizebox{\columnwidth}{!}{
%     \begin{tabular}{rcccccc}
%     \toprule[1.5pt]
%     \multirow{2}{*}{\textbf{Groundtruth}} & \multicolumn{2}{c}{\textbf{CIDEr}} & \multicolumn{2}{c}{\textbf{ROGUE-L}} & \multicolumn{2}{c}{\metricname}  \\ 
%     \cmidrule(lr){2-3}
%     \cmidrule(lr){4-5}
%     \cmidrule(lr){6-7}
%     &\textbf{Pearson} & \textbf{Kendall $\tau$} & \textbf{Pearson} & \textbf{Kendall $\tau$}& \textbf{Pearson} & \textbf{Kendall $\tau$}  \\ \midrule
%     Overall  & 0.84 & 0.37 & 0.57 & 0.32 & -0.21 & -0.19 \\ 
%     % Annotator 2 & & & & &  &\\ 
%     \bottomrule[1.5pt]
%     \end{tabular}
%     }
%     \caption{\small{\textbf{Comparing \modelname against Human Annotations: }We benchmark the performance of \modelname against the annotations obtained from AD experts for $107$ Soccernet video clips.}}
%     \label{tab:human-annotation}
% \end{table}

%  % # ARGE; Kendall Rank correlation: -0.19202; Pearson correlation: -0.21021382614017325
 
%  % # CIDeR: Kendall Rank correlation: 0.37952; Pearson correlation: 0.844451729420619
 
%  %  # ROUGE: Kendall Rank correlation: 0.32362; Pearson correlation: 0.5726759729086413
%%%%%%%%%%%%%%%%%%%%%%%%%
%%%%%%%%%%%%%%%%%%%%%%%%% 

% \subsubsection{Our Metric: Automated Retrieval-based Groundless Evaluation for AD (\metricname)}
\subsubsection{\metricname: Automated Retrieval-based Groundless Evaluation for AD}
\label{subsubsec:arge}

Existing AD evaluation metrics—such as CIDEr, ROUGE, and CRITIC—either depend on reference ADs or capture only surface-level similarity. Yet AD is inherently subjective, allowing multiple valid narrations for the same video. Moreover, these metrics ignore accessibility guidelines. To overcome these limitations, we propose \metricname, a reference-free metric that assesses generated ADs based on compliance with established accessibility standards~\cite{adguide}. These guidelines define rules for high-quality ADs beyond basic narration. We focus on five key aspects: $(i)$ naming characters or players, $(ii)$ maintaining brevity within video duration, $(iii)$ describing major actions or events, $(iv)$ avoiding pronouns, and $(v)$ ensuring descriptive diversity. Traditional word-matching metrics fail to capture such semantic and stylistic nuances, underscoring the need for evaluation methods that recognize equivalent yet distinct descriptions and verify adherence to AD quality standards. We formalize \metricname in Eq.~\ref{eq:argead-eq1}:

% {\small 
% \begin{equation}
%     \text{\metricname} = \frac{1}{N} \sum_{i=1}^{N} \frac{\left( \mathbf{z}_{\mathfrak{p}}^{(i)} + \mathbf{z}_{\mathfrak{a}}^{(i)} + \mathbf{z}_{\mathfrak{l}}^{(i)} + \mathbf{z}_{\mathfrak{pr}}^{(i)} + \mathbf{z}_{\mathfrak{o}}^{(i)} \right)}{5}
%     \label{eq:argead-eq1}
% \end{equation}}

{\small 
\begin{equation}
    \text{\metricname} = \frac{1}{N} \sum_{i=1}^{N} \frac{\left( \mathbf{z}_{{p}}^{(i)} + \mathbf{z}_{{a}}^{(i)} + \mathbf{z}_{{l}}^{(i)} + \mathbf{z}_{{pr}}^{(i)} + \mathbf{z}_{{o}}^{(i)} \right)}{5}
    \label{eq:argead-eq1}
\end{equation}}

\textit{where} $i$ is the video clip index, $N$ is the total number of clips, $\mathbf{z}_p$ is the score for the presence of player names, $\mathbf{z}_a$ for actions/events, $\mathbf{z}l$ for AD length compliance, $\mathbf{z}{pr}$ for discouraging personal pronouns, and $\mathbf{z}_o$ for avoiding overuse or repetition of contextual content.
% =================>
% To address this, we propose a novel evaluation metric,~\metricname, designed to assess ADs according to established AD conventions, providing a robust measure of quality. To begin with, for a given game, we retrieve the player list ${p_1, p_2, \dots, p_N}$ and verify that nouns identified from sentence structure match player names, thus ensuring that ADs consistently refer to players by name. Next, as detailed in Section~\ref{subsec:prompting}, we detect key soccer events and align them with players via frame-level matching. ~\metricname penalizes the use of pronouns by performing grammatical categorization of each word based on sentence structure. Each predicted AD is segmented into parts-of-sentences (POS) using standard NLTK pos-tagger. Based on their POS category the predicted entities are cross-checked with game-specific information retrieved from a structured external database , which we constructed using SoccerNet-v2 annotations, described in Section~\ref{subsec:ad-datasets}. This database contains multiple tables with detailed game information, player data, league context, events, lineups, etc. We make use of basic queries, similar to~\ref{lst:label}, to retrieve the specific information on individual game, league, and season. We formularize the proposed evaluation metric, shown in Eq. \ref{eq:argead-eq1}:

Using the standard NLTK POS tagger, \metricname segments each predicted AD into parts of speech (POS). Words from relevant POS categories are then cross-referenced with game-specific information from an external structured database built using SoccerNet-v2 annotations (Section~\ref{subsec:ad-datasets}). This database includes detailed tables on games, players, leagues, events, and lineups. We use simple queries, as illustrated below, to retrieve information for specific games, leagues, and seasons.

\begin{lstlisting}[basicstyle=\small, label={lst:label},language=SQL]
SELECT game FROM games WHERE season='2014-2015' AND league='england_epl' AND away='Arsenal' AND home='Manchester United'
\end{lstlisting}
In our rule-based analysis, each component of the metric is computed as follows:
\begin{itemize}[noitemsep]
\item Player names ($\mathbf{z}_p^{(i)}$): We build a player index ${p_1, p_2, \dots, p_N}$ from the database and verify that each noun identified by the POS tagger matches a player name, ensuring accurate identification throughout the AD. A score of $1$ is assigned for correct matches, $0$ otherwise.
\item Actions ($\mathbf{z}_a^{(i)}$): Detected verbs are compared against a curated list of top soccer actions (Appendix~\ref{sec:appendix-a}). A score of $1$ is assigned when the correct action is used in context by \textit{AD-VidLlama2}.
\item Pronouns ($\mathbf{z}_{pr}^{(i)}$): To discourage excessive pronoun use, $\mathbf{z}_{pr}^{(i)}$ is set to $0$ if players are referred to only by pronouns.
\item Length compliance ($\mathbf{z}_l^{(i)}$): AD duration is compared to the video length using TTS-generated audio~\cite{10.1109/ICASSP.2018.8461368}. If the AD exceeds the video length, it is penalized; otherwise, $\mathbf{z}_l^{(i)} = 1$.
\item Originality ($\mathbf{z}_o^{(i)}$): To ensure the model does not reuse commentary directly, $\mathbf{z}_o^{(i)}$ is set to $1$ if the Levenshtein\footnote{\url{https://rapidfuzz.github.io/Levenshtein/levenshtein.html}} ratio between the predicted AD and provided commentary is $< 0.5$.
\end{itemize}

Each correctly matched AD rule earns a reward in \metricname, and scores are normalized and averaged across selected AD rules. The \metricname score lie in $[0,1]$, with a score of 1 indicating full compliance with all selected AD guidelines. This evaluation metric is especially useful given that we lack human-annotated ADs for SoccerNet-v2, making \metricname a reliable benchmark for our generated ADs.

\section{Experiments and Results}
\label{sec:experiments}
In this section, we elaborate on the experiments we performed to evaluate~\modelname and validate~\metricname.

%%%%%%%%%%%%%%%%%%%%%%%%%
% \subsection{Comparison of \modelname with Prior Works}
\subsection{Benchmarking \modelname}
\label{subsec:benchmark-evaluation}
% With respect to the domain of sports (soccer games), there is prior work in automating AD and 
We benchmark our results against prior work using the metrics from Section~\ref{subsec:ad-metrics}. As shown in Table~\ref{tab:comparison-movies}, our method matches or outperforms previous approaches on CMD-AD and MAD-Eval. In the sports domain, where no prior AD automation exists, we use SoccerNet video descriptions~(\textit{row 1, Table~\ref{tab:comparison-sports}}) to compute \metricname scores for \textbf{Soccernet-S}. While video captioning and AD generation are related, captions often include excess detail. Despite no domain-specific training, \modelname achieves comparable or better \metricname scores.

\subsection{Qualitative Results}
\label{subsec:qualitative-results}
We show some qualitative results of our approach for both movies and sports in Fig~\ref{fig:qualitative-results-image}.
For each sample in Soccernet-S dataset we show snippets from the game, predicted AD, expert AD annotations, and commentary. Additional soccer AD samples are shown in Appendix \ref{sec:appendix-a}.
\begin{figure*}[t]
    \centering
    \includegraphics[width=0.8\textwidth, scale=0.3]{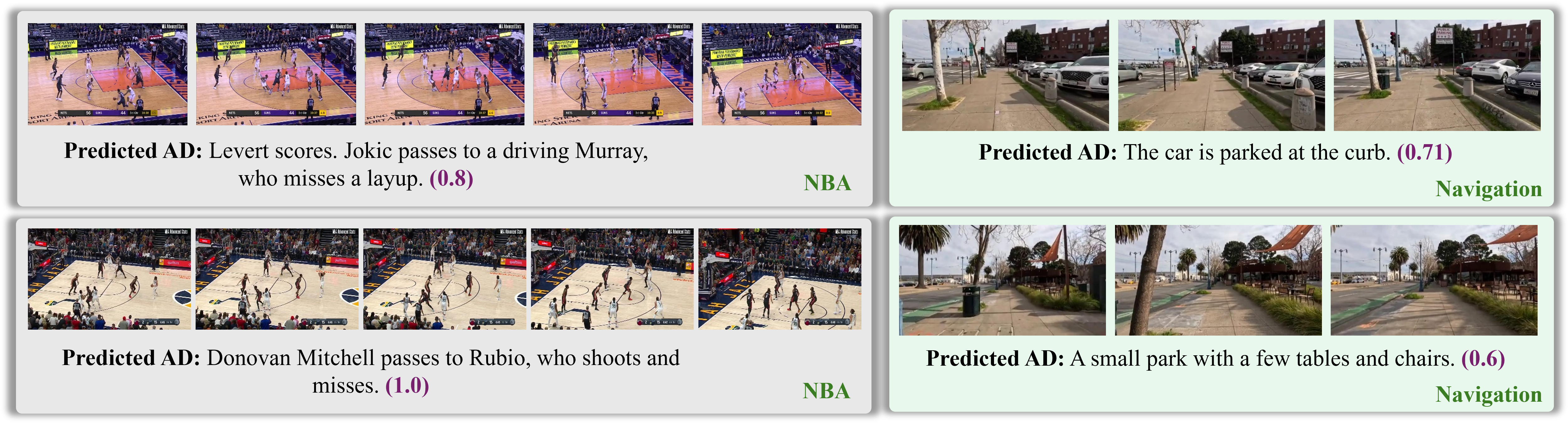}
    \caption{\small{\textbf{Additional Qualitative Results: }We show qualitative visualizations for NBA game (left) and Street Navigation (right).}} 
    \label{fig:qualitative-results-other}
    \vspace{-15pt}
\end{figure*}

\textbf{\ul{In-the-wild videos: }}  We also perform in-the-wild experiments using a random soccer video from YouTube~\footnote{{https://www.youtube.com/watch?v=b9W8sb50BZE}}, which achieved an ARGE-AD score of $0.59$. For this video, we used prior game information available from the YouTube label—teams (Cagliari vs. Fiorentina), season (2014–2015), and league (Italy Serie A). The video, 3.56 minutes long, was segmented into 16 scenes, each processed by \modelname. The \metricname score was averaged across all scenes, with the best performance ($0.59$) obtained when player names and actions were included as contextual cues. Qualitative examples are provided in Appendix Fig.~\ref{fig:in-the-wild-soccer}.

\subsection{Ablation Experiments}
\label{subsec:ablation-expts}
We present additional experiments to evaluate the design and effectiveness of \modelname and \metricname.
\subsubsection{Prompt Ablations for AD Generation}
Choosing an appropriate prompt is critical for achieving strong LLM performance. We therefore ablate the use of different prompts during inference (exact phrasing in Appendix~\ref{sec:appendix-a}). During fine-tuning, we used Prompt~1, a simple one-line instruction. At inference, we tested three strategies: Prompt~1 (basic instruction), Prompt~2 (in-context example), and Prompt~3 (explicitly encoding AD guidelines such as avoiding pronouns). As shown in Table~\ref{tab:prompt-ablations}, performances are largely comparable, but Prompt~3 yields the best results across CMD-AD, MAD-Eval, and \textbf{Soccernet-S}, suggesting that incorporating AD-specific guidelines improves \textit{AD-VidLlama2}’s outputs.

% We present out predicted AD for an NBA game (left) and street navigation scenes (right)

% inferencing with Prompt 2 gives the best results for \textbf{Soccernet-S}. 

% we observe that the example in Prompt 2 does not help, but prompt 3 does help enforce some of the aspects of AD into the AD-VidLlama2 predictions leading to an increased performance.

\subsubsection{Context Ablations for AD Generation}
In Table~\ref{tab:context-ablations}, we modify our prompt $\mathcal{P}$ to include different contextual components—character/player names, actions and events, subtitles/commentary, and the previous clip’s AD (first three columns). We report metrics for the CMD-AD, MAD-Eval, and Soccernet-S datasets. Since subtitles and previous ADs are unavailable for the movie datasets, those components are excluded. Overall, adding context such as character or player names and event details improves performance, while including subtitles/commentary or prior ADs does not consistently yield further gains.

% For uniformity, we consider ARGE-AD scores for Soccernet-S, obtained from prompt 3, similar to the movie dataset.

\subsubsection{Analyzing Human Annotations}
We engaged two AD experts to annotate 107 soccer clips (~10 s each), extracted by’ scene detection \modelname. This study aimed to validate our proposed metric \metricname, similar to previous work~\cite{2023mmvid}. To assess inter-rater reliability, 15 clips (~$20\%$) were annotated by both experts. The metric reliability was evaluated by comparisons with human consensus using CIDEr and ROUGE-L to compute Pearson and Kendall correlations~\cite{graham-etal-2015-accurate} (Table~\ref{tab:human-annotation}). Although Pearson scores showed a significant positive correlation, Kendall values did not. The Intersection-over-union (IoU) between ADs yielded a similarity of $0.19$, indicating notable variability between ADs generated by experts. Glos et al.~\cite{Glos2024-fa} emphasize that, regardless of linguistic style, ADs must objectively capture key visual details as outlined by ACB guidelines~\cite{acb}. To assess this, we computed \metricname scores for AD experts and \modelname. The scores are $0.88$ for expert 1, $0.95$ for expert 2, and $0.73$ for \modelname. These results show that experts closely follow AD guidelines, while \modelname achieves comparable but improvable performance.

\section{Conclusion, Limitations, and Future Scope}
\label{sec:conclusion}
In this work, we propose \modelname, an end-to-end framework for AD generation beyond the movie domain, without requiring human-annotated ADs for training. Focusing on the sports domain, we demonstrate how contextual cues—such as player information, actions, commentary, and prior clips—enrich AD generation. \modelname achieves first-of-its-kind results for soccer games. We also introduce a novel reference-free evaluation metric, \metricname, to assess overall AD quality based on standard AD guidelines. To validate it, we collect expert AD annotations for soccer games, evaluating both \metricname’s effectiveness and how ADs are crafted for fast-paced sports. Qualitative analysis shows that \modelname’s outputs closely match those of human experts (Fig.~3).
Our framework does have limitations that suggest directions for future work. Currently, player identification relies on visual frames, which can fail under occlusion or distant views where jersey numbers are unclear. Incorporating GPS data for player positions and ball tracking could improve context and accuracy. Moreover, while temporal segmentation and ASR enhance contextual grounding, some timing and alignment errors with visual events persist.
% Another important contextual cue can be facial expressions of characters/players involved with the clip. 
In addition, multiple video feeds beyond the final broadcast cut could further enrich the model’s understanding. Future extensions could also personalize ADs based on a viewer’s favorite teams, players, or events. Regarding the evaluation metric, we currently capture only five components from ACB and DCMP guidelines, though many others remain to be incorporated. This work represents an initial step toward developing comprehensive, reference-free metrics for assessing AD quality. As recent approaches, including ours, rely on Video-LLMs that are prone to hallucinations, integrating factual correctness checks will be essential for improving AD reliability.

{\small
\bibliographystyle{references/IEEEtran.bst}
\bibliography{references/egbib.bib}
}

%%%%%%%%% 
% \clearpage
% \input{sections/100_camera_ready}
% \clearpage
%%%%%%%%% 
\clearpage
\appendix
\section{Appendix A}
\label{sec:appendix-a}  
%%%%%%%%%%%%%%%%%%%%%%%%%%%%%%%%%%%%
In this section, we provide more details regarding \modelname and \metricname. 
%%%%%%%%%%%%%%%%%%%%%%%%%%%%%%%%%%%%
% \vspace{-15pt}
\subsection{Summarizing our Algorithm}
% \vspace{-10pt}
\begin{algorithm}[H]
\caption{Generalizing Audio Description Generation}\label{alg:ad-generation}
\begin{algorithmic}[1]
\State \textbf{Input:} $N$ video clips $\mathcal{I}_1, \mathcal{I}_2, \dots, \mathcal{I}_N$
\State \textbf{Output:} Text-based Audio Description (AD) $\mathcal{Y}_{1:N}$

% \State \textbf{\underline{Finetuning the Video-LLM for AD Generation}}
\State \textbf{\underline{Finetuning the Video-LLM for AD Generation}}
    \State Load pre-trained Video-LLM (VideoLlama2)
    \State Finetune the Video-LLM on movie dataset $\mathcal{D} = \left\{\mathcal{I}_{1:M}, \mathcal{Y}_{1:M}\right\}$ with prompt $\mathcal{P}$: ``\textit{Give an audio description of the given video}.''
    \State Minimize loss $\mathcal{L}$ to obtain \textit{AD-VidLlama2} 
    \State Freeze \textit{AD-VidLlama2}

\State \textbf{\underline{Contextual AD Generation with \modelname}}
    \For{each input clip $\mathcal{I}_i$ from $i = 1$ to $N$}
        \State Extract commentary $c_i$ using multi-lingual ASR.
        \State Context retrieval to identify players and actions:
        \State $p_i^j \gets \text{Player Identification}$ using pose keypoint detection and jersey number recognition.
        \State $a_i^k \gets \text{Action Recognition}$ using TAD for soccer events.
        \State Generate AD $\mathcal{Y}_i$
    \EndFor
\end{algorithmic}
\end{algorithm}
\vspace{-10pt}
%%%%%%%%%%%%%%%%%%%%%%%%%%%%%%%%%%%%
% \vspace{-5pt}
\subsection{Prompt Details}
\label{subsec:prompt-details}
\begin{figure}[h]
    \centering
    \includegraphics[width = 0.8\linewidth]{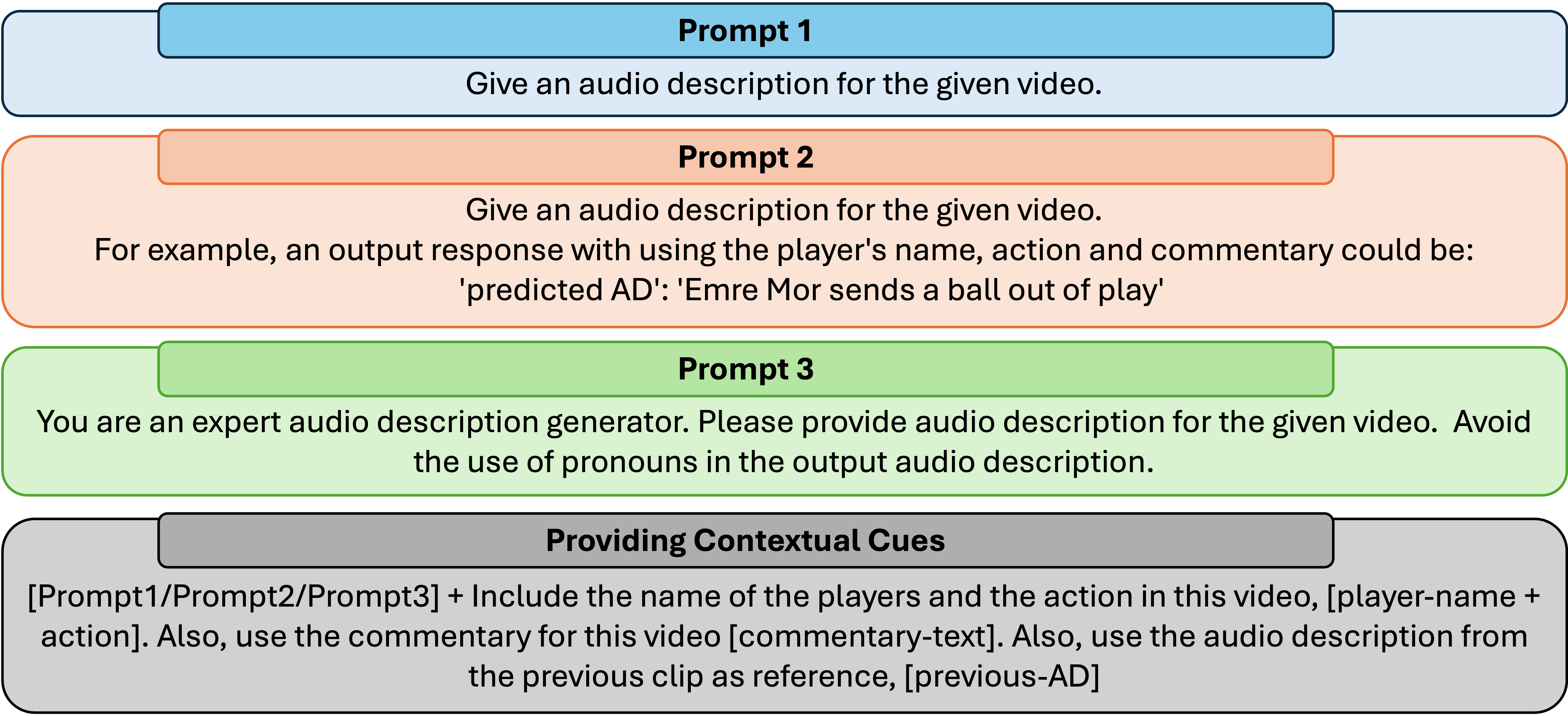}
    \caption{\small{\textbf{Prompting in \modelname: } We present the three prompt variants used to evaluate \modelname. We also depict how we provide contextual cues information along with the prompts.}} 
    % \vspace{-15pt}
    \label{fig:prompt-text-img}
\end{figure}

\modelname uses prompts to query~\textit{AD-VidLlama2} to generate the final AD. We experiment with prompt formats to understanding the effects of prompting on finetuned Video-LLM models. The prompts used in experiments in Tab~\ref{tab:prompt-ablations} are shown in Fig~\ref{fig:prompt-text-img}. Prompt 1 is a one-line instruction; Prompt 2 mirrors in-context learning by adding an example to the prompt, and finally in Prompt 3, we include AD characteristics in the prompt itself. 

% \vspace{-10pt}
\subsection{Soccernet-S Dataset Details}
The original Soccernet-v2 dataset contains $100$ games each with $90$ mins long videos. We segment the videos into scenes using the Temporal Scene Segmentation step in Fig~\ref{fig:overview-inference}\textit{(i)} and create a smaller version of the original dataset. We pick a total of $~2000$ videos to be in our evaluation dataset, \textbf{Soccernet-S}(mall). The dataset contains a variety of scenes, like video moments with major soccer events, zoomed-in scenes focusing on key players, wide view of multiple players, crowd cheering scenes among many more. We will release this list of exact scenes included in the \textbf{Soccernet-S} dataset upon acceptance for future benchmarking. 
%%%%%%%%%%%%%%%%%%%%%%%%%%%%%%%%%%%%
%%%%%%%%%%%%%%%%%%%%%%%%%%%%%%%%%%%%
% \vspace{-15pt}

%%%%%%%%%%%%%%
\begin{figure}[h]
    \centering
    \includegraphics[width = .9\linewidth]{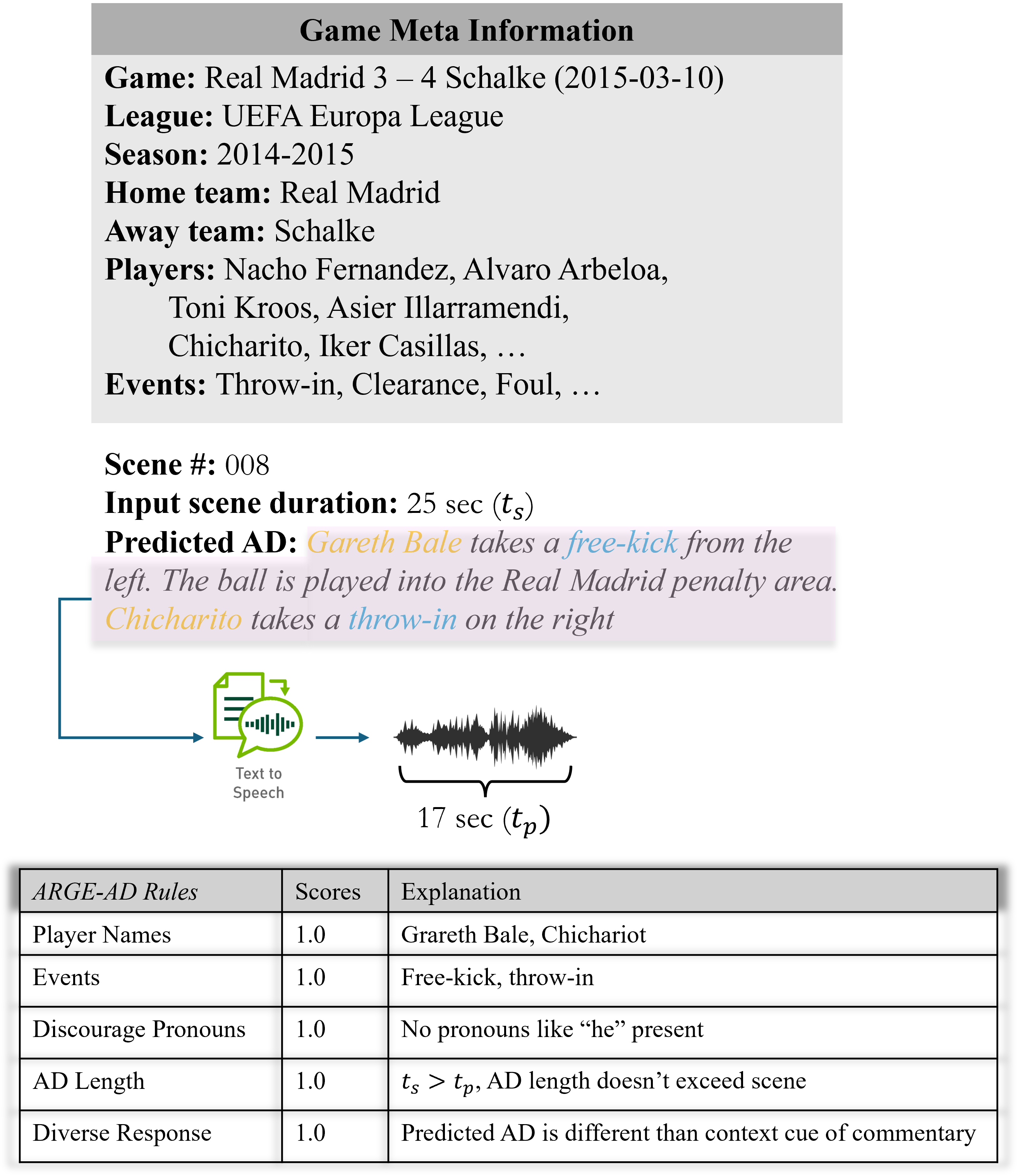}
    \caption{\textbf{\metricname Example Walkthrough: }We present a predicted AD example and walk through scoring mechanism for better understanding of the five components of \metricname.}
    \label{fig:sege-ad-details-process}
\end{figure}
%%%%%%%%%%%

\begin{table*}[t]
    \centering
    \resizebox{.8\textwidth}{!}{
    \begin{tabular}{cccccccccc}
        \toprule[1.5pt]
         \multirow{2}{*}{\textbf{\large \# frames}}& \multicolumn{5}{c}{\textbf{\large CMD-AD}} & \multicolumn{3}{c}{\textbf{\large MAD-Eval}}  & \textbf{\large Soceernet-S} \\
         \cmidrule(rr){2-6}
         \cmidrule(lr){7-9}
         \cmidrule(ll){10-10}
         &  \textbf{CIDEr} & \textbf{ROUGE-L} & \textbf{CRITIC} & \textbf{LLM-AD-EVAL}  & \textbf{\metricname} & \textbf{CIDEr} & \textbf{ROUGE-L} & \textbf{\metricname} & \textbf{\metricname} \\
        \midrule
        \Large \textbf{$4$}  & 9.30 & 18.17 & 9.36 & 0.92 & 39.97 & 10.12 & 18.49 & 0.41 & 0.73\\
        % \midrule
        \Large \textbf{$8$}  & 10.83 & 18.43 & 6.72 & 1.07 & 38.85 & 11.52 & 18.78 & 0.39 & 0.67\\
        % \midrul
        \Large \textbf{$16$}  & \textbf{15.15} & \textbf{19.35} & \textbf{16.16} & 1.28 & 38.99 & 12.32 & 18.83 & 0.40 & 0.43\\
        \bottomrule[1.5pt]
    \end{tabular}}
    \caption{\textbf{Ablations Experiments for \# Frames: }We perform an ablation experiment with \textit{AD-VidLlama2} for $4$, $8$ and $16$ frames of the input videos; and test it for the three settings as well for both the movie datasets, CMD-AD and MAD-Eval and also the SoccerNet dataset.}
        \label{tab:numframes-ablations}
\end{table*}

\begin{figure*}[h]
    \centering
    \includegraphics[width =0.95\textwidth, scale=0.3]{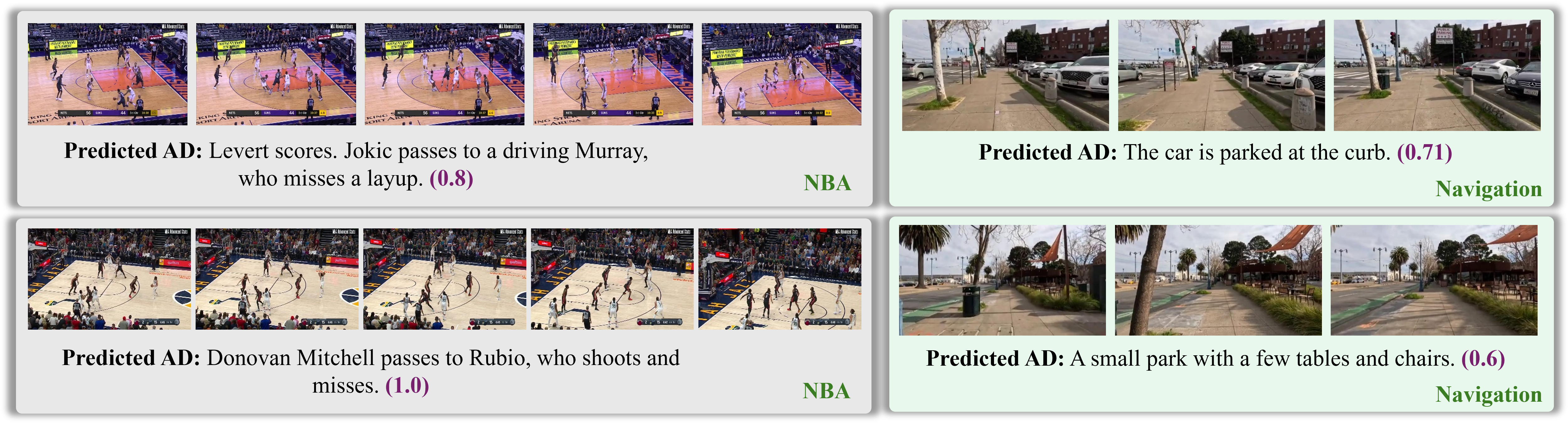}
    \caption{\small{\textbf{Additional Qualitative Results: }We show qualitative visualizations for NBA game (left) and Street Navigation (right).}} 
    \label{fig:qualitative-results-other}
    %\vspace{-15pt}
\end{figure*}
%%%%%%%%%%%%%%%%%%%%%%%%%%%%%%%%%%%%

\subsection{Additional \textbf{\metricname} Details}
\label{subsec:arge-details}
Due to the lack readily available ground-truths, we introduce \metricname. We visualize the process of computing ARGE-AD scores for each sample. Each predicted AD has a game meta information associated with it, which acts as a reference labels against which entities like player names are compared. The length of an AD shouldn't exceed the original video/scene, we employ off-the-shelf text-to-speech models to aid us in comparing the duration of scenes and the predicted AD text. Each of these rules are awarded a score between 0 and 1. We explain the metric pictorially in Fig~\ref{fig:sege-ad-details-process}.

\textbf {{How can \metricname be improved? }} We believe that the proposed metric, \metricname is just a first attempt towards the need of a reference-free metric grounded on AD rules and characteristics. However, we only incorporate five such rules as laid down by the American Council of Blind. We believe there is a lot more work that can be done to enhance this metric. For instance, currently our metric does not check for factual correctness of character/player names, actions and other context. This becomes more important with increased use of LLMs and VLMs for the task, given their hallucinating nature. We also do not check for temporal fluency and redundancy of the predicted ADs. It is important to ensure that while the AD generation happens per video clip, it makes coherent sense and not be repetitive. Furthermore, currently we weight the five AD characteristics equally in the computation of \metricname. This is a research question to understand which rules play a more significant role in assessing the AD quality.

% \subsection{Our Research Work's Ethical Considerations} 
% This Automated AD generation approach is developed to enhance accessibility. Although we aim to produce accurate and fair descriptions, our resulting ADs may not always fully capture the context or intent of the scene. If deployed, we will continually monitor the system and also encourage feedback for improvement from our users. Our goal is to provide inclusive content while respecting privacy and diverse perspectives. 
% \clearpage
\section{Appendix B}
\label{sec:appendix-b}  
\begin{figure*}[h]
    \centering
    \includegraphics[width = 0.85\linewidth]{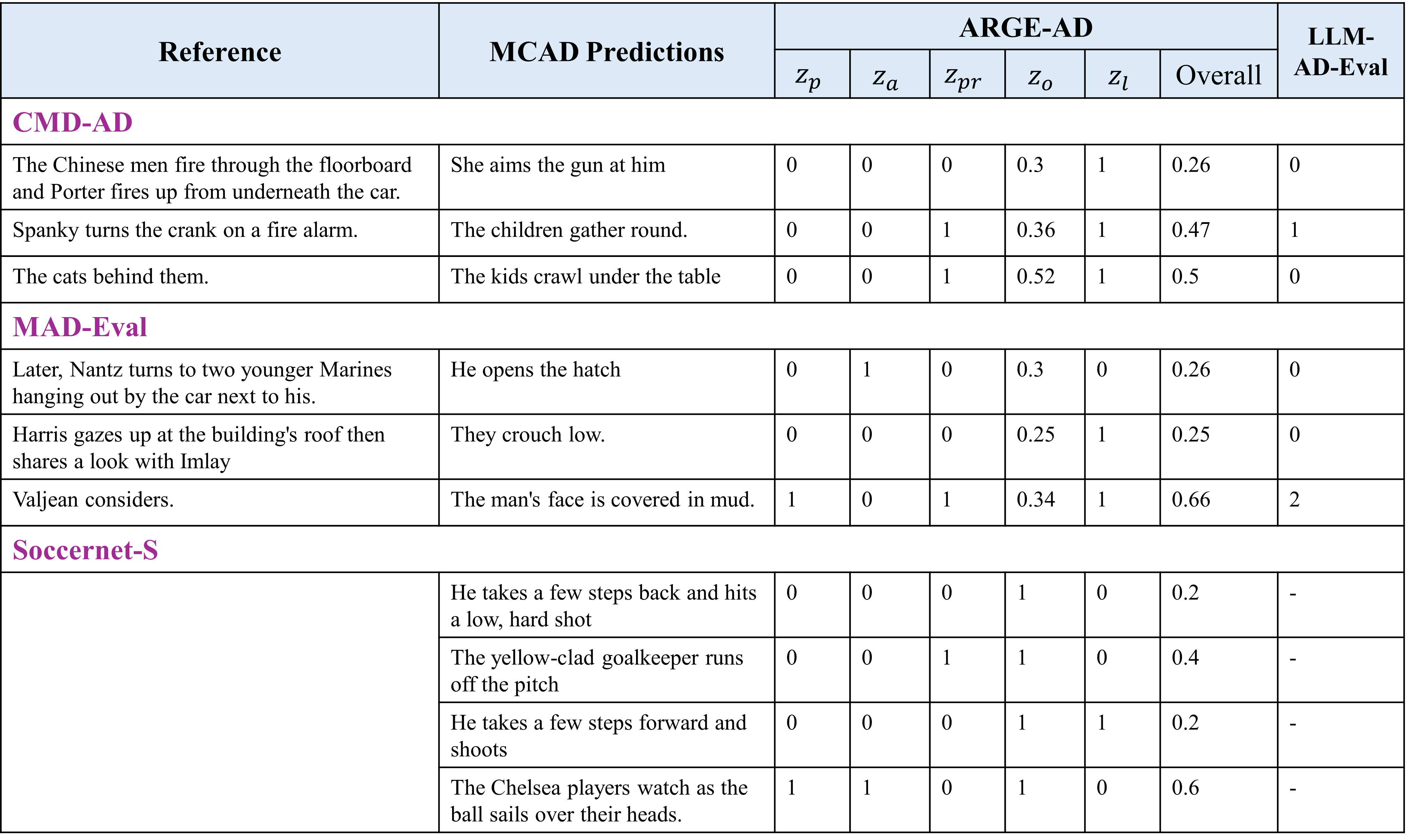}
    \caption{\small{\textbf{Some Failure Cases~(CMD-AD, MAD-eval and SoccerNet-S Dataset): }Examples where \modelname predictions are not the best. We show \metricname scores and LLM-AD-Eval scores for each \modelname predicted AD.}}
    \label{fig:bad-examples}
\end{figure*}

In this section, we provide more results for more ablation experiments and qualitative results.
%\vspace{-15pt}

\subsection{\textbf{\metricname} Analysis}
\label{subsec:arge-analysis}
To further investigate the \metricname, in Fig~\ref{fig:arge-ad-img}, we plot the score values of the five components separately for the different context settings~(Tab~\ref{tab:context-ablations}, cols 1-3). We see that that adding playername and actions to the context in the prompt increases the values of $\mathbf{z}_{\mathfrak{p}}$ and $\mathbf{z}_{\mathfrak{a}}$ respectively, but we notice that adding commentary and prior clip's AD hurts $\mathbf{z}_{\mathfrak{pr}}$ and  $\mathbf{z}_{\mathfrak{l}}$ as the length of the AD increases, mimicking the commentary; so that more pronouns are present in the AD. The plot indicates that \metricname is indeed evaluating the AD generations on these various aspects. 
%\vspace{-15pt}
\subsection{Additional Ablation Experiments}
\label{subsec:additional-ablation} 
\subsubsection{AD Generation w/o Finetuning}
\label{subsubsec:ad-gen-no-finetuning}
\begin{figure}[h]
    \centering
    \includegraphics[width = .95\linewidth]{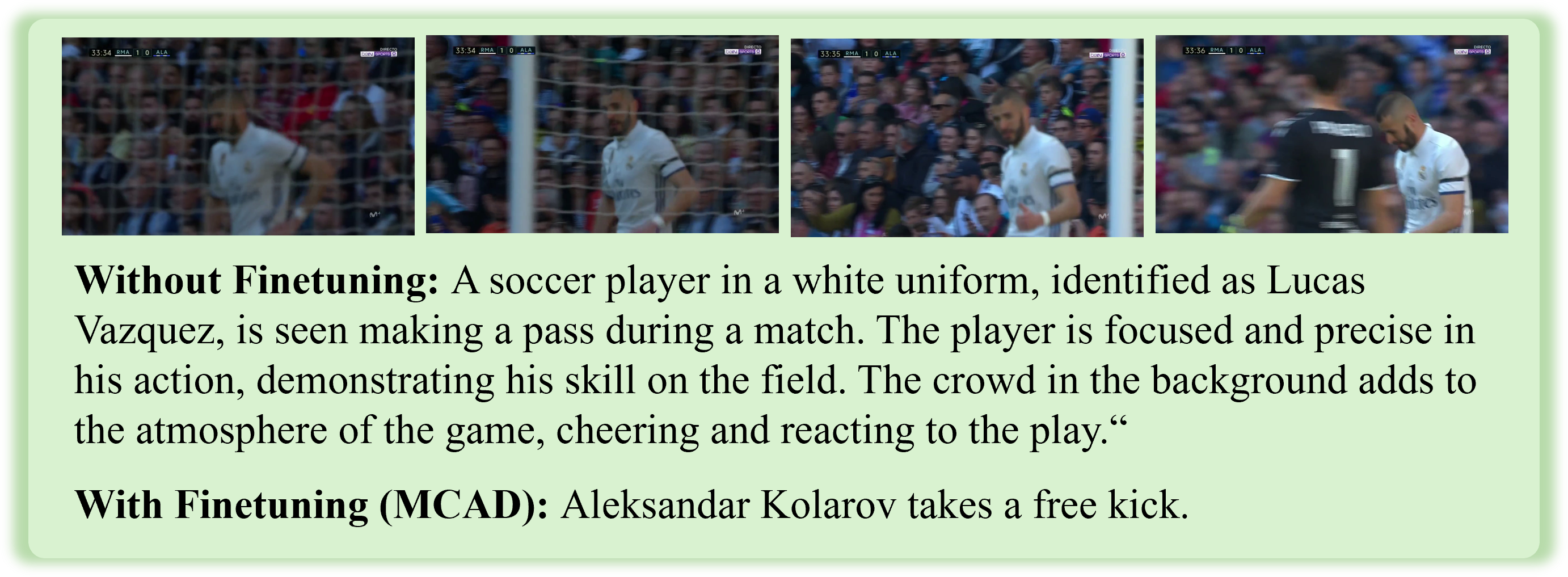}
    \caption{\small{\textbf{Impact of finetuning Video-LLM: }With this qualitative example we show the importance of finetuning the Video-LLM for predicting AD.}}
    \label{fig:finetune}
    %\vspace{-20pt}
\end{figure}

\begin{figure}[h]
    \centering
    \includegraphics[width=.7\linewidth]{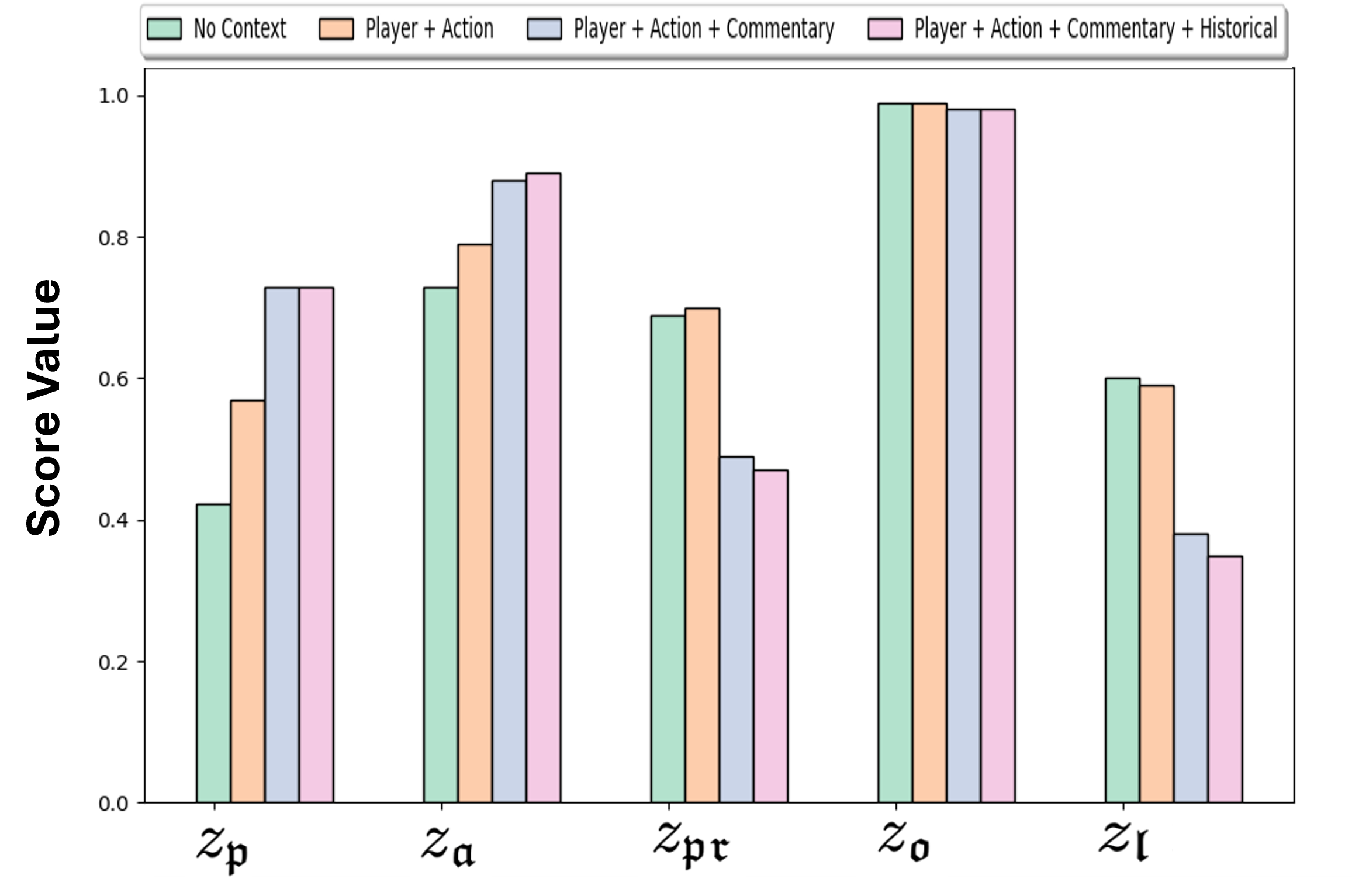}
    \caption{\small{\textbf{Investigating the five \metricname Components: } We plot the scores of the individual components of \metricname for the varying levels of context information in the prompt.}}
    \label{fig:arge-ad-img}
\end{figure}

We experiment generation of AD with and without finetuning. We get an \metricname of $0.27$ for CMD-AD-Eval which drops significantly without finetuning the Video-LLM. The Soccernet-S dataset majorly is penalized for the length of the generated AD, as shown in Fig~\ref{fig:finetune}.

\subsubsection{\# Frames Ablations for AD Generation}
\label{subsubsec:frames-ablation4}
We experiment with the number of frames during finetuning and inference to understand the performance when the frame sampling value is changed. We finetune the model for three different frame sampling frequency, $4$, $8$, and, $16$ and then use these finetuned versions for inference. We have summarized our results in Tab~\ref{tab:numframes-ablations}. While prior works~\cite{han2023autoad, zhang2024mm, wang2024contextual} propose setting the number of frames to $8$, our experiments show that using a larger number of frames, like $16$, can produce better results. 

%%%%%%%%%%%%%%%%%%%%%%%%%%%%%%%%%%%%%
%%%%%%%%%%%%%%%%%%%%%%%%%%%%%%%%%%%%%
%%%%%%%%%%%%%%%%%%%%%%%%%%%%%%%%%%%%%
%%%%%%%%%%%%%%%%%%%%%%%%%%%%%%%%%%%%%

%%%%%%%%%%%%%%%%%%%%%%%%%%%%%%%%%%%%
\begin{figure}[h]
    \centering
    \includegraphics[width = \linewidth]{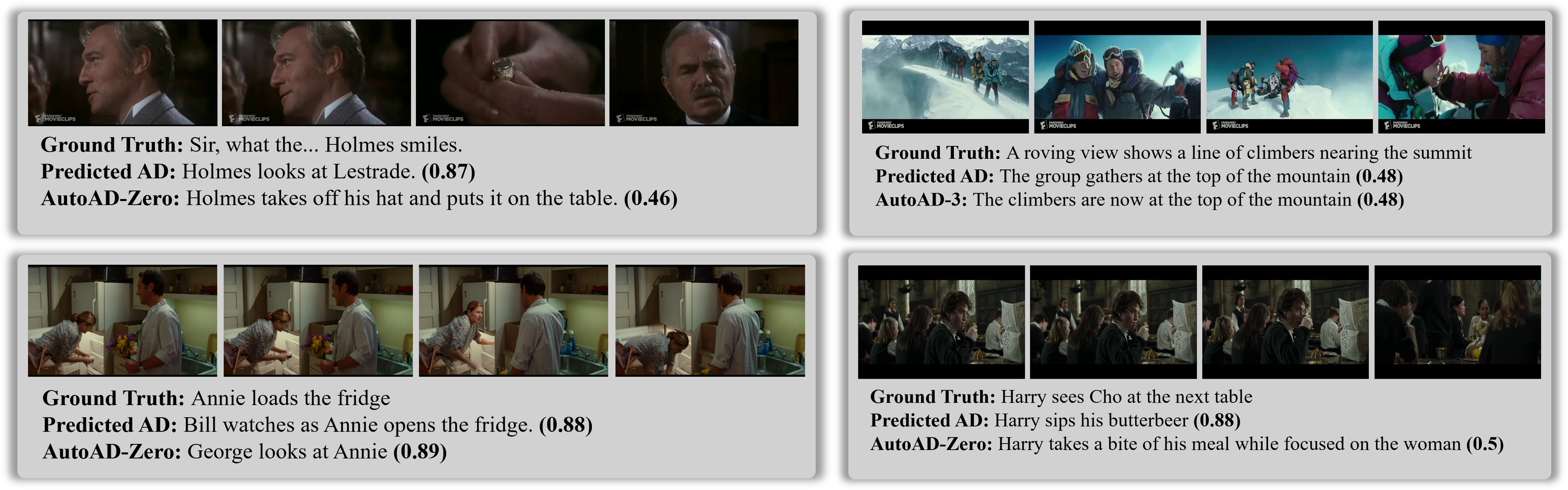}
    \caption{\small{\textbf{Additional Qualitative Results (CMD-AD and MAD-eval Dataset):} We demonstrate additional qualitative results on CMD-AD \textit{(top row)} and MAD-eval \textit{(bottom row)}.}}
    \label{fig:additional-qual-supp-movies}
\end{figure}
%%%%%%%%%%%%%%%%%%%%%%%%%%%%%%%%%%%%
\begin{figure}[h]
    \centering
    \includegraphics[width = \linewidth]{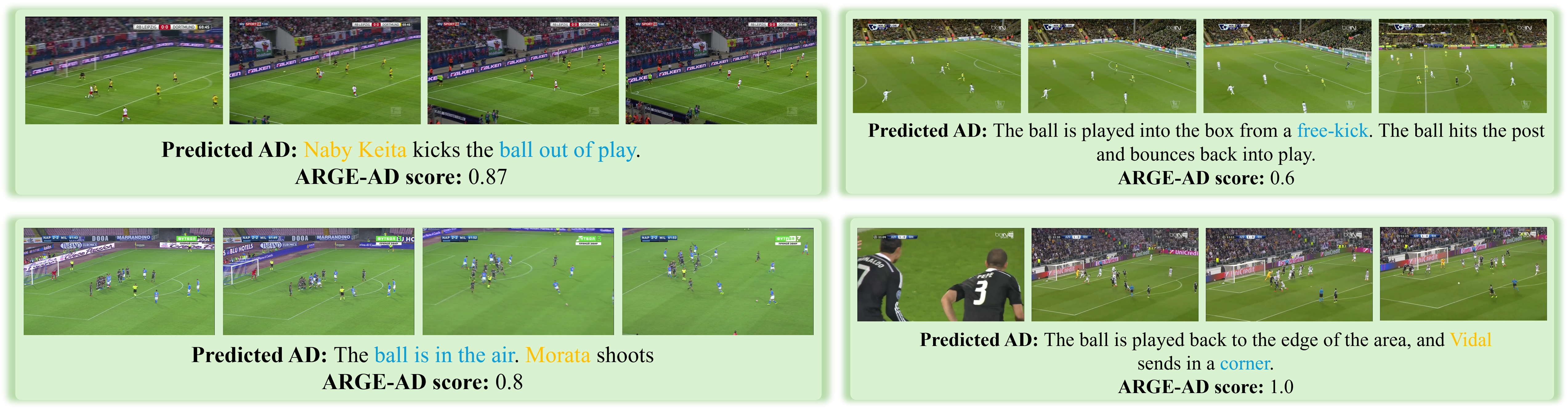}
    \caption{\small{\textbf{Additional Qualitative Results (SoccerNet Dataset):} We select samples from different soccer leagues. The samples are from \textit{Norwich vs Chelsea [England EPL (2015--2016)]}, \textit{Juventus vs Real Madrid [UEFA Champions League (2014--2015)]}, \textit{RB Leipzig vs Dortmund [Germany Bundesliga (2016--2017)]}, and \textit{Real Madrid vs Betis [Spain LaLiga (2015--2016)]} (from top).}}
\label{fig:additional-qual-supp-soccernet}
\end{figure}
%%%%%%%%%%%%%%%%%%%%%%%%%%%%%%%%%%%%%
\begin{figure}[h]
    \centering
    \includegraphics[width = 0.85\linewidth]{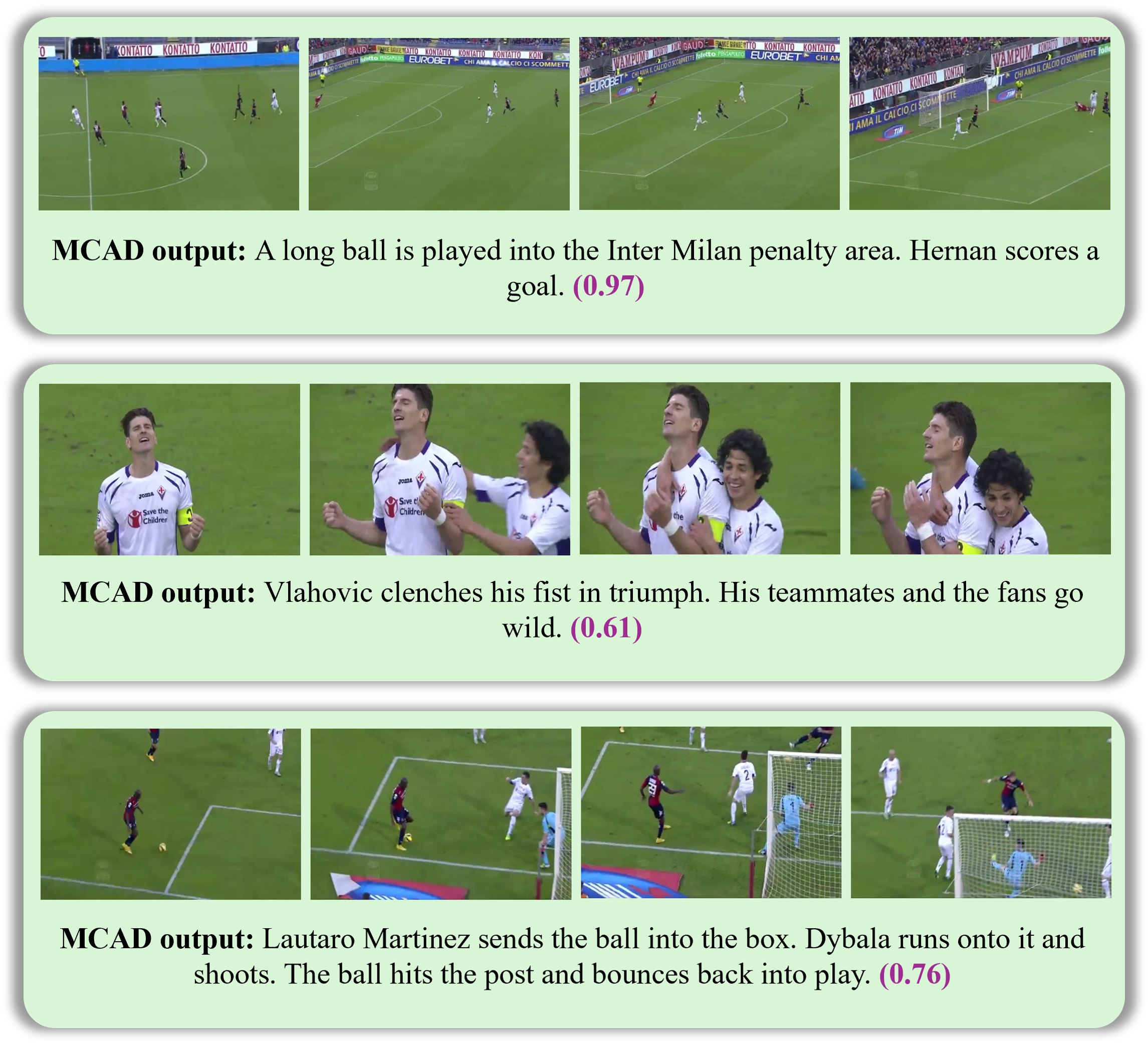}
    \caption{\small{\textbf{Qualitative Results for In-the-Wild Soccer Videos: } We show results on three clips extracted from the YouTube video and the corresponding \metricname score.}}
    \label{fig:in-the-wild-soccer}
\end{figure}
%%%%%%%%%%%%%%%%%%%%%%%%%%%%%%%%%%%%
%%%%%%%%%%%%%%%%%%%%%%%%%%%%%%%%%%%%

%\vspace{-15pt}
\subsection{Additional Qualitative Results}
\begin{enumerate}
    \item We present some more qualitative results of \modelname, for each of the movie dataset~(Fig~\ref{fig:additional-qual-supp-movies}) and also SocccerNet-S dataset~(Fig~\ref{fig:additional-qual-supp-soccernet}). The ARGE-AD scores for \modelname gives higher scores due to the presence of character name and shorter AD length, shown in 2nd example in Fig \ref{fig:additional-qual-supp-movies}. Fig \ref{fig:additional-qual-supp-movies}, 3rd example~(show similar \metricname for both \modelname and AutoAD-III predictions. We present some failure cases in Fig \ref{fig:bad-examples}. 
    \item We also add some qualitative results for in-the-wild AD generation by \modelname on soccer videos extracted from YouTube in Fig~\ref{fig:in-the-wild-soccer}.
    \item As mentioned with simple modifications to the context that we can retrieve, we are able to extend \modelname for a different sport and also different applications beyond movies and sports. Towards the same, we show some preliminary results of \modelname's AD generations for basketball and also for the application of real-life navigation. We use NSVA~\cite{dew2022sports} dataset to generate AD for basketball. We used their initial annotations to obtain the player name and actions occurring in the video. For the experiment, we used \~20 videos and predicted their corresponding AD by passing the videos and context information through our pipeline. The resultant~\metricname score of $0.77$ is averaged over the these $20$ input videos~(average duration 10 secs). We show some qualitative results for these in Fig~\ref{fig:qualitative-results-other}~\textit{(left)}. To show the generalizability of our \modelname, we also deploy our approach for AD generation for real-life street navigation. We selected a random video\footnote{https://www.youtube.com/watch?v=PiYCBVfxLy8} from YouTube~(resulting in 16 scenes with average duration of 11 seconds). To extract contextual cues, we used of-the-shelf object detection algorithm~\cite{ultralytics2025} and daily action list from human action recognition datasets. The~\metricname scores for navigation use-case is $0.64$. Qualitative results for this are shown in Fig \ref{fig:qualitative-results-other}~\textit{(right)}. 
\end{enumerate}

\subsection{Extending \modelname To Other Domains}
\label{subsec:extension}

We believe our framework can be applied to other sports and also various domains beyond movies and sports, such as navigation, theater, and interactive museum experiences, with only minor modifications and without the explicit need for human-created ADs. To this end, we show the robustness of~\modelname and~\metricname for another sport, basketball and also for navigation. 

We use NSVA~\cite{dew2022sports} dataset to generate AD for basketball. We used their initial annotations to obtain the player name and actions occurring in the video. For the experiment, we used \~20 videos and predicted their corresponding AD by passing the videos and context information through our pipeline. The resultant~\metricname score of $0.77$ is averaged over the these $20$ input videos~(average duration 10 secs). We show some qualitative results for these in Fig~\ref{fig:qualitative-results-other}~\textit{(left)}. To show the generalizability of our \modelname, we also deploy our approach for AD generation for real-life street navigation. We selected a random video\footnote{https://www.youtube.com/watch?v=PiYCBVfxLy8} from YouTube~(resulting in 16 scenes with average duration of 11 seconds). To extract contextual cues, we used of-the-shelf object detection algorithm~\cite{ultralytics2025} and daily action list from human action recognition datasets. The~\metricname scores for navigation use-case is $0.64$. Qualitative results for this are shown in Fig \ref{fig:qualitative-results-other}~\textit{(right)}.

\end{document}